\documentclass[12pt,a4paper]{article}

\usepackage{ifthen} 
\usepackage{subfigure}
\newboolean{pdflatex}
\setboolean{pdflatex}{true} 

\newboolean{articletitles}
\setboolean{articletitles}{true} 

\newboolean{uprightparticles}
\setboolean{uprightparticles}{false} 

\usepackage{microtype}
\usepackage{lineno}  
\usepackage{xspace} 

\usepackage{graphicx}  
\usepackage{color}
\usepackage{colortbl}
\graphicspath{{./figs/}} 

\usepackage{amsmath} 
\usepackage{amssymb}
\usepackage{amsfonts}
\usepackage{upgreek} 

\newcommand*\patchAmsMathEnvironmentForLineno[1]{%
\expandafter\let\csname old#1\expandafter\endcsname\csname #1\endcsname
\expandafter\let\csname oldend#1\expandafter\endcsname\csname
end#1\endcsname
 \renewenvironment{#1}%
   {\linenomath\csname old#1\endcsname}%
   {\csname oldend#1\endcsname\endlinenomath}%
}
\newcommand*\patchBothAmsMathEnvironmentsForLineno[1]{%
  \patchAmsMathEnvironmentForLineno{#1}%
  \patchAmsMathEnvironmentForLineno{#1*}%
}
\AtBeginDocument{%
\patchBothAmsMathEnvironmentsForLineno{equation}%
\patchBothAmsMathEnvironmentsForLineno{align}%
\patchBothAmsMathEnvironmentsForLineno{flalign}%
\patchBothAmsMathEnvironmentsForLineno{alignat}%
\patchBothAmsMathEnvironmentsForLineno{gather}%
\patchBothAmsMathEnvironmentsForLineno{multline}%
}

\usepackage{hyperref}    
\usepackage[all]{hypcap} 

\def\ux85 {\mbox{UX85}\xspace}

\ifthenelse{\boolean{uprightparticles}}%
{

 \def\Pmu         {\ensuremath{\upmu}\xspace}

 \def\Ppi         {\ensuremath{\uppi}\xspace}

 \def\PDelta      {\ensuremath{\Delta}\xspace}                 
 \def\PXi      {\ensuremath{\Xi}\xspace}                 
 \def\PLambda      {\ensuremath{\Lambda}\xspace}                 
 \def\PSigma      {\ensuremath{\Sigma}\xspace}                 
 \def\POmega      {\ensuremath{\Omega}\xspace}                 
 \def\PUpsilon      {\ensuremath{\Upsilon}\xspace}                 
 
 \def\PB      {\ensuremath{\mathrm{B}}\xspace}                 
                  
 \def\PD      {\ensuremath{\mathrm{D}}\xspace}

 \def\PK      {\ensuremath{\mathrm{K}}\xspace}

 \def\Pi      {\ensuremath{\mathrm{i}}\xspace}

}
{

 \def\Pmu         {\ensuremath{\mu}\xspace}

 \def\Ppi         {\ensuremath{\pi}\xspace}

 \mathchardef\PDelta="7101
 \mathchardef\PXi="7104
 \mathchardef\PLambda="7103
 \mathchardef\PSigma="7106
 \mathchardef\POmega="710A
 \mathchardef\PUpsilon="7107
                  
 \def\PB      {\ensuremath{B}\xspace}                 
                  
 \def\PD      {\ensuremath{D}\xspace}

 \def\PK      {\ensuremath{K}\xspace}

 \def\Pi      {\ensuremath{i}\xspace}

}

\def\mup        {\ensuremath{\Pmu^+}\xspace}
\def\mun        {\ensuremath{\Pmu^-}\xspace} 
\def\mumu       {\ensuremath{\Pmu^+\Pmu^-}\xspace}

\def\pion  {\ensuremath{\Ppi}\xspace}

\def\pip   {\ensuremath{\pion^+}\xspace}

\def\kaon  {\ensuremath{\PK}\xspace}
  \def\Kbar  {\kern 0.2em\overline{\kern -0.2em \PK}{}\xspace}

\def\Kz    {\ensuremath{\kaon^0}\xspace}
\def\Kzb   {\ensuremath{\Kbar^0}\xspace}
\def\KzKzb {\ensuremath{\Kz \kern -0.16em \Kzb}\xspace}
\def\Kp    {\ensuremath{\kaon^+}\xspace}
\def\Km    {\ensuremath{\kaon^-}\xspace}

\def\KpKm  {\ensuremath{\Kp \kern -0.16em \Km}\xspace}

\def\Kstarz  {\ensuremath{\kaon^{*0}}\xspace}
\def\Kstarzb {\ensuremath{\Kbar^{*0}}\xspace}

  \def\Dbar    {\kern 0.2em\overline{\kern -0.2em \PD}{}\xspace}
\def\D       {\ensuremath{\PD}\xspace}

\def\Dz      {\ensuremath{\D^0}\xspace}
\def\Dzb     {\ensuremath{\Dbar^0}\xspace}
\def\DzDzb   {\ensuremath{\Dz {\kern -0.16em \Dzb}}\xspace}
\def\Dp      {\ensuremath{\D^+}\xspace}
\def\Dm      {\ensuremath{\D^-}\xspace}

\def\DpDm    {\ensuremath{\Dp {\kern -0.16em \Dm}}\xspace}

\def\B       {\ensuremath{\PB}\xspace}
\def\Bbar    {\ensuremath{\kern 0.18em\overline{\kern -0.18em \PB}{}}\xspace}

\def\Bz      {\ensuremath{\B^0}\xspace}
\def\Bzb     {\ensuremath{\Bbar^0}\xspace}

\def\Bdb     {\ensuremath{\Bbar^0}\xspace}

  \def\Y#1S{\ensuremath{\PUpsilon{(#1S)}}\xspace}

\def\Lbar {\ensuremath{\kern 0.1em\overline{\kern -0.1em\PLambda}}\xspace}

\newcommand{\decay}[2]{\ensuremath{#1\!\to #2}\xspace}         

\def\to                 {\ensuremath{\rightarrow}\xspace}

\def\qsq       {\ensuremath{q^2}\xspace}

\def\CP                {\ensuremath{C\!P}\xspace}

\def\AT#1     {\ensuremath{A_{\mathrm{T}}^{#1}}\xspace}           

\def\C#1      {\ensuremath{\mathcal{C}_{#1}}\xspace}                       
\def\Cp#1     {\ensuremath{\mathcal{C}_{#1}^{'}}\xspace}                    
\def\Ceff#1   {\ensuremath{\mathcal{C}_{#1}^{\mathrm{(eff)}}}\xspace}        
\def\Cpeff#1  {\ensuremath{\mathcal{C}_{#1}^{'\mathrm{(eff)}}}\xspace}       
\def\Ope#1    {\ensuremath{\mathcal{O}_{#1}}\xspace}                       
\def\Opep#1   {\ensuremath{\mathcal{O}_{#1}^{'}}\xspace}                    

\newcommand{\tev}{\ensuremath{\mathrm{\,Te\kern -0.1em V}}\xspace}
\newcommand{\gev}{\ensuremath{\mathrm{\,Ge\kern -0.1em V}}\xspace}
\newcommand{\mev}{\ensuremath{\mathrm{\,Me\kern -0.1em V}}\xspace}
\newcommand{\kev}{\ensuremath{\mathrm{\,ke\kern -0.1em V}}\xspace}
\newcommand{\ev}{\ensuremath{\mathrm{\,e\kern -0.1em V}}\xspace}
\newcommand{\gevc}{\ensuremath{{\mathrm{\,Ge\kern -0.1em V\!/}c}}\xspace}
\newcommand{\mevc}{\ensuremath{{\mathrm{\,Me\kern -0.1em V\!/}c}}\xspace}
\newcommand{\gevcc}{\ensuremath{{\mathrm{\,Ge\kern -0.1em V\!/}c^2}}\xspace}
\newcommand{\gevgevcccc}{\ensuremath{{\mathrm{\,Ge\kern -0.1em V^2\!/}c^4}}\xspace}
\newcommand{\mevcc}{\ensuremath{{\mathrm{\,Me\kern -0.1em V\!/}c^2}}\xspace}

\def\deriv {\ensuremath{\mathrm{d}}}

\def\gsim{{~\raise.15em\hbox{$>$}\kern-.85em
          \lower.35em\hbox{$\sim$}~}\xspace}
\def\lsim{{~\raise.15em\hbox{$<$}\kern-.85em
          \lower.35em\hbox{$\sim$}~}\xspace}

\def\tell1  {TELL1\xspace}
\def\ukl1   {UKL1\xspace}

\usepackage{cite} 
\usepackage{mciteplus}

\usepackage{longtable} 
\usepackage{caption}

\def\BdbKstbMuMu     {\decay{\Bdb}{\Kstarzb \mup\mun}}

\usepackage{xcolor}

\begin{document}
\renewcommand{\thefootnote}{\fnsymbol{footnote}}
\setcounter{footnote}{1}

\begin{titlepage}

{\bf\boldmath\Large
\begin{center}
  Method for an unbinned measurement of the $q^2$ dependent decay amplitudes of
  \BdbKstbMuMu decays
\end{center}
}

\vspace*{2.0cm}

\begin{center}
U.~Egede$^1$, M.~Patel$^1$,  K.A.~Petridis$^{2,1}$ \bigskip\\
{\it\footnotesize 
$ ^1$Imperial College London, London, United Kingdom \\ \vspace{0.2cm}
$ ^2$Univeristy of Bristol, Bristol, United Kingdom \\ \vspace{0.2cm}
}
\end{center}

\vspace{\fill}

\begin{abstract}
  \noindent
  A method for determining the $q^2$ dependent $\Kstarzb$ spin
  amplitudes of \BdbKstbMuMu decays through a maximum likelihood fit
  to data is presented. While current experimental techniques extract
  a limited set of observables in bins of $q^2$, our approach allows
  for the determination of all observable quantities as continuous
  distributions in $q^2$. By doing this, the method eliminates the
  need to correct theory predictions of these observables for $q^2$
  averaging effects, thus increasing the sensitivity to the effects of
  physics beyond the Standard Model. Accounting for the symmetries of
  the angular distribution and using a three parameter ansatz for the
  $q^2$ dependence of the amplitudes, the precision of the angular
  observables and the sensitivity to new physics is estimated using
  simulated events. These studies are based on the sample sizes
  collected by the LHCb experiment during Run-I and expected for
  Run-II.
\end{abstract}

\vspace*{2.0cm}
\vspace{\fill}

\end{titlepage}

\pagestyle{empty}  


\newpage
\setcounter{page}{2}
\mbox{~}

\cleardoublepage


\renewcommand{\thefootnote}{\arabic{footnote}}
\setcounter{footnote}{0}
\tableofcontents
\cleardoublepage


\pagestyle{plain} 
\setcounter{page}{1}
\pagenumbering{arabic}


%

\section{Introduction}\label{sec:Introduction}

Rare $b\to s\mumu$ processes are suppressed in the Standard Model (SM)
as they can proceed only via electroweak penguin or box type diagrams.
As-yet undiscovered particles could give additional contributions with
comparable amplitudes to those of the SM processes, and such decays
are therefore sensitive probes of new phenomena.  

The angular distribution of the $K^{-}\pi^{+}\mumu$ system in
\decay{\Bzb}{\Kstarzb\mumu} decays is of particular interest, as it
can be described by a number of measurable quantities which are
sensitive to new physics and can be precisely predicted in a given
physics model. Theoretical predictions for such observables are
particularly precise in the range of dimuon invariant mass squared,
$q^2$,
$1<q^2<6~\gevgevcccc$~\cite{Altmannshofer:2008dz,Bobeth:2008ij,Bobeth:2010wg,Das:2012kz,Horgan:2013pva,
  Mahmoudi:2014mja,Hambrock:2013zya,Hurth:2014vma}. The potential of
the \decay{\Bzb}{\Kstarzb\mumu} decay as a probe of New Physics (NP)
has resulted in numerous proposals for observables with varying levels
of theoretical precision, as described in
Refs.~\cite{Kruger:1999xa,Kruger:2005ep,ref:egede_matias_reece,Altmannshofer:2008dz,Bobeth:2011gi,DescotesGenon:2012zf,Descotes-Genon:2013vna}.

The dominant uncertainty in the predictions of
\decay{\Bzb}{\Kstarzb\mumu} observables, is attributed to the
calculation of the $\Bzb\to\Kstarzb$ hadronic form-factors.  The LHCb
collaboration have determined a number of observables which are
designed to have a reduced dependence on these
form-factors~\cite{LHCb-PAPER-2013-037,LHCb-CONF-2015-002}. The
measurement of the observable $P_{5}'$~\cite{DescotesGenon:2012zf},
which is designed to ensure the cancellation of the hadronic
$\Bzb\to \Kstarzb$ soft form-factors at leading order, exhibits a
local tension at the level of 3.7~$\sigma$ with respect to the SM
prediction of Ref~\cite{DescotesGenon:2012zf}.  This measurement has
been interpreted as an indication of a new, heavy, vector
particle~\cite{Descotes-Genon:2013wba,Altmannshofer:2013foa,Beaujean:2013soa,Hurth:2013ssa,Jager:2012uw,
  Jager:2014rwa,Descotes-Genon:2014uoa,Altmannshofer:2014cfa,Crivellin:2015mga,
  Gauld:2013qja,Altmannshofer:2014rta,Mahmoudi:2014mja, Datta:2013kja,
  Sierra:2015fma} or as a consequence of previously unaccounted for
QCD effects~\cite{Lyon:2014hpa,Khodjamirian:2012rm,Khodjamirian:charmloop}.
In addition to potentially unexpected QCD effects, there are several
other factors that limit the sensitivity of the data to new physics
effects. As detailed below, these include the omission of certain
symmetry relations between the observables in experimental analyses,
the interference between P- and S-waves of the $K^-\pi^+$ system, and
the binning in $q^2$.

The LHCb collaboration have published a number of separate analyses of
\decay{\Bzb}{\Kstarzb\mumu} decays to determine different
observables~\cite{LHCb-PAPER-2013-019,LHCb-PAPER-2013-037}.  Recently,
the LHCb collaboration also presented a measurement of all observables
related to the $K^-\pi^+$ system in a P-wave
state~\cite{LHCb-CONF-2015-002}. In all these measurements, the
simple relations between the P-wave observables, arising due to the
symmetries of the angular distribution, are implemented. However, the
remaining complex relation between P-wave
observables~\cite{ref:egede_matias_reece,ref:quim_anatomy}, is not
exploited. This results in redundant parameters being determined,
reducing the overall experimental precision of the observables.

One of the dominant systematic uncertainties for the experimental
determination of \decay{\Bzb}{\Kstarzb\mumu} observables is the lack
of knowledge of the S-wave contribution to the predominately P-wave
$K^{-}\pi^{+}$ system. An S-wave contribution can induce a bias on the
(P-wave) observables and dilute the experimental sensitivity to such
observables~\cite{ref:blake_egede_shires_swave}. Until recently, the
experimental results included a systematic uncertainty to cover
this. In the latest LHCb analysis~\cite{LHCb-CONF-2015-002}, the
S-wave components are explicitly included. However, further symmetry
relations between the P- and S-wave components are not accounted for,
resulting in two redundant S-wave
parameters~\cite{quim_swave_sym}. As in the case of the omitted
P-wave symmetry relations, this redundancy dilutes the experimental
precision of the observables of the angular distribution.

The precision of global fits to existing measurements of
\decay{\Bzb}{\Kstarzb\mumu} transitions are also limited by the fact
that the measurements are performed in bins in $q^2$.  Comparison
between theoretical predictions and experimental measurements then
requires the integration of theoretically predicted observables over
experimental $q^2$ bins. This integration has the effect of diluting
variations and introducing dependence on other, potentially more
poorly predicted, observables. For example, the angular terms in the
differential decay rate that involve the theoretically clean $P_{5}'$
observable have a $\sqrt{F_{\rm L}(1-F_{\rm L})}$ prefactor, where
$F_{\rm L}$ is the longitudinal polarisation fraction of the \Kstarzb
and is a poorly predicted observable. This results in the experimental
measurements being sensitive to,
$\int{\sqrt{F_{\rm L}(q^2)(1-F_{\rm L}(q^2))}} P_{5}'(q^2) dq^2$.
Although the angular distribution is also sensitive to
$\int{F_{\rm L}(q^2) dq^2}$, this does not enable
$\int{ P_{5}'(q^2) dq^2}$ to be computed and hence does not allow the
full exploitation of the cancellation of the form factors at leading
order for which $P_5'$ was designed\footnote{Experimental
  measurements~\cite{LHCb-CONF-2015-002,LHCb-PAPER-2013-019,LHCb-PAPER-2013-037}
  also enable the $q^2$-averaged $P_5'$ to be computed using the ratio
  $\langle S_{5}\rangle/\sqrt{\langle F_{\rm L}\rangle(1-\langle F_{\rm L}\rangle)}$, where
  $\langle S_5\rangle$ and $\langle F_{\rm L}\rangle$ denote $q^2$-averaged
  quantities. This ratio is not the same as the optimal
  observable $\langle S_5/\sqrt{F_L(1-F_{\rm L})}\rangle$.}.

In this paper, we propose a method of analysing
\decay{\Bzb}{\Kstarzb\mumu} decays that allows the determination of
all of the \Kstarzb amplitudes as a parametric function of $q^2$. This
method allows the formation of any observable from a single fit to the
data from a given experiment, including the full experimental
correlations. The method also allows the S-wave related amplitudes to
be determined, removing the need for any experimental systematic
uncertainty from S-wave contamination.  The $q^2$ shape information
and the application of all the symmetry relations of the angular
distribution, result in a substantial gain in sensitivity to new
physics effects and completely remove the $q^2$ averaging problem
mentioned above. The method can be used with the
\decay{\Bzb}{\Kstarzb\mumu} sample that is already available at the
LHCb experiment.
  
  The paper is organised as follows: Sec.~\ref{sec:diffdecayrate}
  describes the decay rate and angular observables of
  $\decay{\Bzb}{K^-\pi^+\mumu}$ transitions;
  Sec.~\ref{sec:amplitudes_method} describes the method for extracting
  the \Kstarzb helicity amplitudes from a single fit to data;
  Sec.~\ref{sec:amplitudes_results} presents the results of the method
  using simulated \decay{\Bzb}{\Kstarzb\mumu} decays with sample sizes
  equivalent to those obtained by LHCb during Run-I and projected for
  Run-II of the LHC; finally Sec.~\ref{sec:amplitudes_comparison}
  compares the sensitivity of the amplitude fit to other methods for
  extracting the \Kstarzb angular observables.

\section{The differential decay rate}\label{sec:decayrate}

\label{sec:diffdecayrate}
The differential decay rate of the \Bzb meson, for a $K\pi$ system
in a P-wave configuration and ignoring scalar contributions to the
dimuon system, is given by~\cite{Kruger:1999xa}

\begin{align}
\frac{\deriv^{4}\Gamma[\decay{\Bzb}{\Kstarzb\mumu}]}{\deriv\cos\theta_{\ell}\,\deriv\cos\theta_{K}\,\deriv\phi\,\deriv q^{2}} 
= & \frac{9}{32\pi} [\frac{}{}  {J_{1s}} \sin^{2} \theta_{K} + {J_{1c}}\cos^{2}\theta_{K}  + {J_{2s}}\sin^{2}\theta_{K} \cos 2\theta_{\ell}  ~+ \nonumber\\
   &  \phantom{\frac{9}{32\pi}}~\frac{}{}{J_{2c}} \cos^{2} \theta_{K} \cos 2\theta_{\ell} + {J_{3}} \sin^{2}\theta_{K}\sin^{2} \theta_{\ell} \cos 2\phi ~+\nonumber\\
   &\phantom{\frac{9}{32\pi}}~\frac{}{} {J_{4}} \sin 2\theta_{K} \sin2\theta_{\ell} \cos\phi +  {J_{5}} \sin 2\theta_{K} \sin\theta_{\ell}\cos\phi ~+\label{eq:decrate}\\
   & \phantom{\frac{9}{32\pi}}~\frac{}{}{J_{6s}} \sin^{2}\theta_{K}\cos\theta_{\ell} + {J_{7}}  \sin 2\theta_{K} \sin\theta_{\ell} \sin\phi ~+\nonumber\\
   & \phantom{\frac{9}{32\pi}}~\frac{}{}{J_{8}} \sin 2\theta_{K} \sin2\theta_{\ell}\sin\phi +  {J_{9}} \sin^{2} \theta_{K} \sin^{2}\theta_{\ell} \sin 2\phi\frac{}{} ]\nonumber.
\end{align} 

\noindent The angular observables, $J_i(q^2)$, depend on
six \qsq dependent complex amplitudes, $A_0^{L,R}$,
$A_{\parallel}^{L,R}$, $A_{\perp}^{L,R}$ representing the three
polarisation states of the \Kstarzb. The configuration of a longitudinally
polarised \Kstarz and time-like polarised dimuon system is suppressed
and therefore safely neglected. The labels $L$ and $R$ refer to
the chirality of the dimuon system. The various $J_i(q^2)$
observables are given by

\begin{equation}
\label{eq:masterEqJs1}
\begin{split}
J_{1s} &= \frac{(2+\beta_\mu^2)}{4} \bigl[ |A_\perp^L|^2 + |A_\parallel^L|^2 + (L\to R) \bigr] + \frac{4m_\mu^2}{q^2} \mathrm{Re}(A_\perp^L A_\perp^{R*} + A_\parallel^L A_\parallel^{R*})\\
J_{1c} &= |A_0^L|^2 + |A_0^R|^2 + \frac{4m_\mu^2}{q^2} \bigl[2\mathrm{Re}(A_0^L A_0^{R*}) \bigr]\\
J_{2s} &= \frac{\beta_\mu^2}{4}\biggl[ |A_\perp^L|^2 + |A_\parallel^L|^2 + (L\to R)\biggr]\\
J_{2c} &= -\beta_\mu^2\biggl[ |A_0^L|^2 + (L\to R)\biggr]\\
J_3 &= \frac{\beta_\mu^2}{2}\biggl[ |A_\perp^L|^2 - |A_\parallel^L|^2 +(L\to R)\biggr]\\
J_4 &= \frac{\beta_\mu^2}{\sqrt{2}}\biggl[ \mathrm{Re}( A_0^LA_\parallel^{L*} ) +(L\to R)\biggr]\\
J_5 &= \sqrt{2}\beta_\mu\biggl[ \mathrm{Re}( A_0^LA_\perp^{L*} ) -(L\to R)\biggr]\\
J_{6s} &= 2\beta_\mu\biggl[ \mathrm{Re}( A_\parallel^LA_\perp^{L*} ) -(L\to
R)\biggr]\\
J_7 &= \sqrt{2}\beta_\mu \biggl[ \mathrm{Im} ( A_0^LA_\parallel^{L*} ) -(L\to R)\biggr]\\
J_8 &= \frac{\beta_\mu^2}{\sqrt{2}}\biggl[ \mathrm{Im} ( A_0^LA_\perp^{L*} ) +(L\to R)\biggr]\\
J_9 &= \beta_\mu^2 \biggl[ \mathrm{Im} ( A_\parallel^{L*}A_\perp^{L} ) +(L\to R)\biggr]
\end{split}
\end{equation}

\noindent with $\beta_\mu^{2} = (1 - 4m_{\mu}^{2}/\qsq)$ and
$(L\to R)$ denotes the same term but with flipped chirality.  In the
limit of $\qsq \gg 4 m_{\mu}^{2}$ the various $J_i$ coefficients can
be trivially related by $J_{2c}=-J_{1c}$ and $J_{2s}=J_{1s}/3$.  An
additional, more complicated relation exists between $J_{2c}$ and the
rest of the angular observables as noted in
Ref.~\cite{ref:egede_matias_reece} and explicitly given in
Ref.~\cite{ref:quim_anatomy}.

While the differential decay rate in Eq.~\ref{eq:decrate} is defined
for the decay of the $\Bzb$ meson, the decay of the $\Bz$ can be given in
complete analogy, by starting from Eq.~\ref{eq:decrate} and performing
the substitution $J_{i}\to \bar{J}_{i}$, following the angular convention
described in Refs.~\cite{LHCb-PAPER-2013-019,LHCb-PAPER-2013-037}
This convention allows the sum of the decay rates
of \Bz and \Bzb mesons to be written in terms of sums of $J_i$ and
$\bar{J}_i$ angular observables by simply performing the substitutions
$J_{i}\to (\bar{J}_{i}+J_{i})$ into Eq.~(\ref{eq:decrate}). It is
therefore convenient to define \CP-averaged observables $S_i$, as
discussed in Ref.~\cite{Altmannshofer:2008dz},

\begin{align}
S_i&\equiv\frac{J_i+\bar{J}_i}{\left({\rm d}\Gamma+{\rm d}\bar{\Gamma}\right)/{\rm d}q^2}.
\end{align}

\noindent In a similar way, the \CP-asymmetric observables, $A_i$, are defined as~\cite{Altmannshofer:2008dz}
\begin{align}
A_i&\equiv\frac{J_i-\bar{J}_i}{\left({\rm d}\Gamma+{\rm d}\bar{\Gamma}\right)/{\rm d}q^2}.
\end{align}

\subsection{S-wave interference} 
\label{sec:Swave}
The expression in Eq.~(\ref{eq:decrate}) assumes that the $\Km\pip$ system is in a P-wave
configuration, as is the case for the $\Kstarz(892)$ vector
meson. To account for a $\Km\pip$ system in an S-wave configuration,
the spin-amplitudes need to be modified to account for the presence of
the S-wave amplitudes $A_{00}^{L,R}$.

In previous experimental analyses~\cite{LHCb-PAPER-2013-019,LHCb-PAPER-2013-037} the presence
of an S-wave contribution was accounted for by assigning a systematic
uncertainty. The method presented here enables the determination of the
S-wave amplitudes using a modified version of Eq.~\ref{eq:decrate},

\begin{equation}
\label{eq:decrateswave}
\begin{split}
\frac{\deriv^{4}\Gamma}{\deriv\cos\theta_{\ell}\,\deriv\cos\theta_{K}\,\deriv\phi\,\deriv  q^{2}} ~\to~ &
\frac{\deriv^{4}\Gamma}{\deriv\cos\theta_{\ell}\,\deriv\cos\theta_{K}\,\deriv\phi\,\deriv
  q^{2}}~+\\
\frac{9}{32\pi} [~{J'}_{1c} (1-\cos 2\theta_{\ell})~+ & {J''}_{1c}\cos\theta_{K}(1-\cos 2\theta_{\ell} ) ~+\\
\phantom{\frac{9}{32\pi}[}~{J'}_{4} \sin 2\theta_{\ell} \sin\theta_{K} \cos\phi ~+            & {J'}_{5} \sin \theta_{\ell} \sin\theta_{K} \cos\phi~+\\
\phantom{\frac{9}{32\pi}[}~{J'}_{7} \sin\theta_{\ell} \sin\theta_{K} \sin\phi~+                 & {J'}_{8} \sin 2\theta_{\ell} \sin\theta_{K} \sin\phi~],
\end{split} 
\end{equation}

\noindent with

\begin{equation}
\label{eq:masterEqJs2}
\begin{split}
{J'}_{1c} &= \frac{1}{3}|A_{00}^L|^2 + \frac{1}{3}|A_{00}^R|^2 \\
{J''}_{1c} &= \frac{2}{\sqrt{3}} \biggl[\mathrm{Re}( A_{00}^L A_0^{L*}) + ( L \to R )\biggr]  \\
{J'}_4 & = \sqrt{\frac{2}{3}}  \biggl[ \mathrm{Re}( A_{00}^L A_{\parallel}^{L*}) + ( L \to R )\biggr]  \\
{J'}_5 & = 2 \sqrt{\frac{2}{3}}  \biggl[ \mathrm{Re}( A_{00}^L A_{\perp}^{L*}) - ( L \to R )\biggr]  \\
{J'}_7 & = 2 \sqrt{\frac{2}{3}}  \biggl[ \mathrm{Im}( A_{00}^L A_{\parallel}^{L*}) - ( L \to R )\biggr]  \\
{J'}_8 & = \sqrt{\frac{2}{3}}  \biggl[ \mathrm{Im}( A_{00}^L A_{\perp}^{L*}) + ( L \to R )\biggr],  \\
\end{split} 
\end{equation}
\noindent as given in Ref.~\cite{Descotes-Genon:2013vna},
where an implicit integration over the mass of the
$K^-\pi^+$ system is assumed. 

The fraction of the S-wave contribution is defined as:
\begin{equation}
F_{\rm S}(q^2) = \frac{|A_{00}^{L}|^{2}+|A_{00}^{R}|^{2} }{\deriv\Gamma/\deriv\qsq} 
\end{equation} 
Where $\deriv\Gamma/\deriv\qsq$ is defined as the total differential
rate of both S and P-wave contributions, given by
\begin{equation}
\begin{split}
\frac{\deriv\Gamma}{\deriv\qsq} & = \frac{\deriv\Gamma_{\rm
    S}}{\deriv\qsq}  + \frac{\deriv\Gamma_{\rm P}}{\deriv\qsq} \\ 
& =|A_{00}^{L}|^{2} +  |A_{0}^{L}|^{2}  +  |A_{\parallel}^{L}|^{2} +  |A_{\perp}^{L}|^{2}   + ( L \to R ). \\ 
\end{split}
\end{equation}

\section{Fitting for the ${\boldsymbol{K^{*0}}}$  amplitudes}
\label{sec:amplitudes_method}
\subsection{Infinitesimal symmetries of the angular distribution}
\label{sec:amplitudes_inf_sym}
Ignoring the S-wave terms, the angular distribution of
\decay{\Bzb}{\Kstarzb\mumu} decays can be described by eleven angular
observables ($J_i$) for each \Bz flavour. These observables are made
up of bilinear combinations of the $K^{*0}$ spin amplitudes and
represent the ``experimental'' degrees of freedom. If the $J_i$ terms
are all independent, the experimental degrees of freedom should match
the number of amplitude components which represent the
``theoretical'' degrees of freedom. However, there are continuous
symmetry transformations of the amplitudes that leave the decay rate
invariant~\cite{ref:egede_matias_reece}. In order for the degrees of freedom
to match it is required that

\begin{equation}
\label{eq:dof_counting}
n_j-n_d=2n_a-n_s,
\end{equation}

\noindent where $n_j$ is the number of $J_i$ terms, $n_d$ the number
of the relations between the $J_i$, $n_a$ is the number of complex
amplitude components and $n_s$ is a number of continuous symmetry
transformations of the amplitudes that leave the decay rate invariant.
In the massless limit $(\qsq \gg 4 m_{\mu}^{2})$, and ignoring scalar
contributions to the dimuon system, there are four continuous symmetry
transformations of the amplitudes ($n_s=4$) that leave each of the
$J_i$, and therefore the decay rate invariant, (see
\cite{ref:egede_matias_reece,ref:quim_anatomy} for a detailed
discussion). Given $n_j=11$ and $n_a=12$, as
specified by Eq.~\ref{eq:dof_counting}, there are three relations
between the various $J_i$, yielding eight independent angular
observables.

The following continuous transformations of the amplitudes, leave the
angular distribution unchanged~\cite{ref:egede_matias_reece,ref:quim_anatomy}:

\begin{equation} \label{eq:symm_transf}n'_i = \left(
    \begin{array}{cc} \textcolor{black}{e^{i\phi_{L}}} & 0 \\ 0 &
      \textcolor{black}{e^{-i\phi_{R}} } \end{array} \right) \left(
    \textcolor{black}{\begin{array}{cc} \cos\theta &
        -\sin\theta \\ \sin\theta & \cos\theta \end{array} } \right)
  \left( \textcolor{black}{ \begin{array}{cc} \cosh i\omega & -\sinh
        i\omega \\ -\sinh i\omega & \cosh i\omega \end{array} }
  \right) n_i~, 
\end{equation}

\noindent  where the basis vectors $n_i$ are defined as,

 \begin{equation}
n_{\parallel} = \left( \begin{array}{c} A_{\parallel}^{L} \\ A_{\parallel}^{R*} \end{array} \right) ~,~ n_{\perp} = \left( \begin{array}{c} A_{\perp}^{L} \\ -A_{\perp}^{R*} \end{array} \right) ~,~n_{0} =\left( \begin{array}{c} A_{0}^{L} \\ A_{0}^{R*} \end{array} \right),
\end{equation}

\noindent The components $\phi_{L}$ and $\phi_{R}$ are phase-rotations of the left- and
right-handed amplitudes separately. The second and third matrices
act as a transformation between the left- and right-handed amplitudes.

The angular distribution is degenerate under these transformations of
the amplitudes. A likelihood function including all twelve real
amplitude components, would therefore exhibit a 4D hypersurface of
continuous maxima in amplitude space, rendering useless the
minimisation techniques for the determination of the amplitude
components.  The symmetries of the angular distribution allow for the
transformation of the amplitudes to a particular basis, where four of
the amplitude components are fixed to some arbitrary value at every
point in $q^2$. The choice of the basis, referred to as
``basis-fixing'', lifts the degeneracy.  For the basis-fixing to allow
for a subsequent fit of the remaining eight amplitudes as a function
of $q^2$, it is required that the values for $\phi_L$, $\phi_R$,
$\theta$, and $\omega$ exist for every point in $q^2$; and the
amplitudes in this transformed basis are slowly varying in $q^2$, such
that they can be described by a simple functional form. This second
requirement restricts the $q^2$ range where the amplitudes can be
extracted. The presence of potential light resonances below
$\sim 1$~\gevgevcccc and of $c\bar{c}$ resonances above 8~\gevgevcccc
motivates the use of the resonance-free and theoretically preferred
region of $1<q^2<6$~\gevgevcccc.

\begin{figure}[!htb]
\centering
\includegraphics[width=0.69\textwidth]{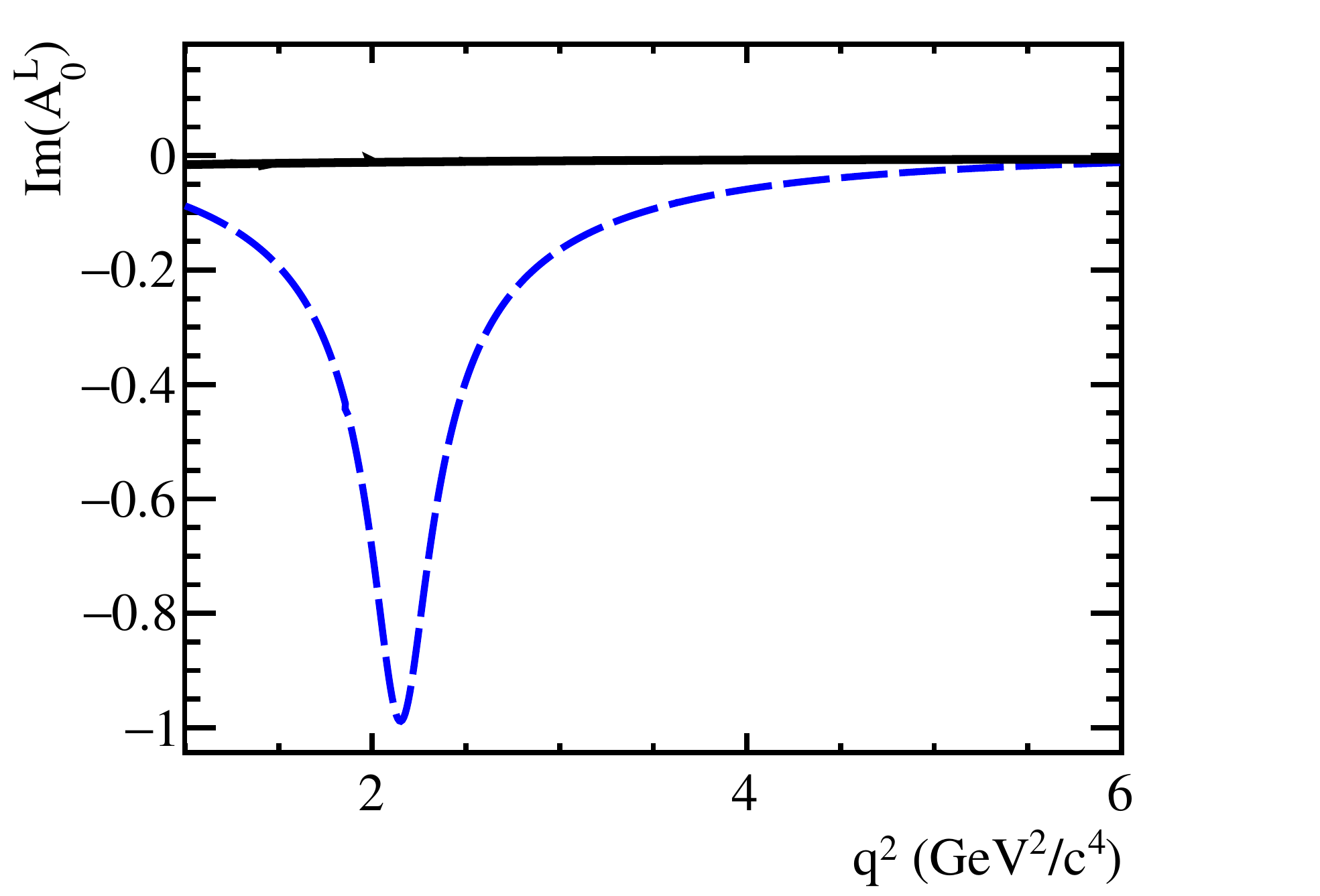}
\caption{ {\small Distribution of the SM $\mathrm{Im}(A_{0}^{L})$
    amplitude in the untransformed (solid black) and in the
    fixed-basis proposed in Ref.~\cite{ref:egede_matias_reece} (dashed
    blue). The untransformed amplitudes are given by the \texttt{EOS}
    program~\cite{Bobeth:2010wg}, up to an overall scaling factor.
    \label{fig:disc_zr_l_im}
  }}
\end{figure}

A previous study described in Ref.~\cite{ref:egede_matias_reece} used the
following basis-fixing
\begin{equation}
  \mathrm{Re}(A_{\parallel}^{L})=\mathrm{Im}(A_{\parallel}^{L})=\mathrm{Im}(A_{\parallel}^{R})=\mathrm{Im}(A_{\perp}^{R})=0.
  \label{eqn:bad_cnstr}
\end{equation}
This basis suffers from a rapidly varying behaviour in $\mathrm{Im}(A_{0}^{L})$ at
$q^2\sim 2\gevgevcccc$, as shown in Fig.~\ref{fig:disc_zr_l_im}. In
Ref.~\cite{ref:egede_matias_reece}, the problems caused by this
discontinuity were avoided by ignoring the $q^2$ region below
2.5\gevgevcccc. 

\begin{figure}[!htb]
\centering
\includegraphics[width=0.75\textwidth]{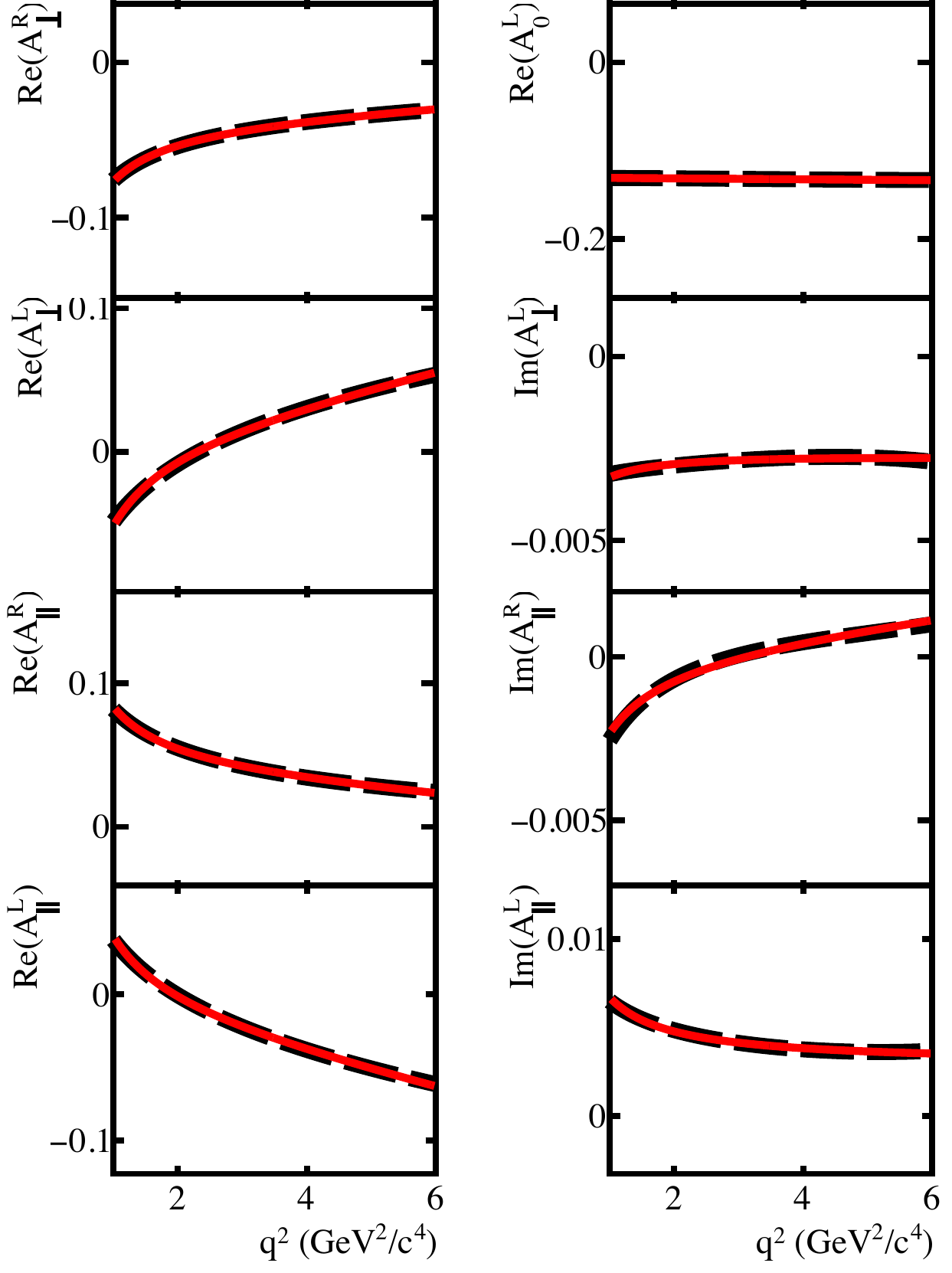}
\caption{ {\small Distribution of the transformed SM $\Kstarz(892)$
    spin amplitudes in the fixed-basis (dashed black line) which
    exhibit a smooth behaviour in $q^2$. The solid red line denotes
    the result of the fit of the $q^2$ dependent ansatz discussed in
    the Sec.~\ref{sec:acc_sym}. The untransformed amplitudes are given
    by the \texttt{EOS} program~\cite{Bobeth:2010wg}, up to an overall
    scaling factor.  Only the non-zero amplitude components in the
    fixed-basis are shown.}
\label{fig:param_SM}
}
\end{figure}

Ignoring the $q^2$ region below 2.5\gevgevcccc is clearly highly
undesirable. For the method that we propose here, we instead use,
\begin{equation}
  \mathrm{Re}(A_{0}^{R})=\mathrm{Im}(A_{0}^{R})=\mathrm{Im}(A_{0}^{L})=\mathrm{Im}(A_{\perp}^{R})=0.
\label{eqn:good_cnstr}
\end{equation}
The amplitudes in the improved fixed-basis then exhibit a slow varying
behaviour in $q^2$ both in the SM, shown in Fig.~\ref{fig:param_SM} as
well as in a range of new physics models. The \texttt{EOS}
program~\cite{Bobeth:2010wg} is used to generate the amplitudes in the
original basis.

\subsection{Exact discrete symmetries}
\label{sec:disc_sym}
In addition to the continuous transformations of the amplitudes that
leave the angular distribution invariant, the angular distribution is
also invariant under discrete transformations of the amplitudes. Even
after the basis-fixing, which reduces the number of amplitudes to
eight, there is a discrete symmetry comprising a simultaneous shift
$A_i \to -A_i$ for all $i$, that leaves the angular distribution
invariant. This symmetry can be seen simply by inspecting
Eq.~\ref{eq:masterEqJs1} and noting that even after the conditions of
Eq.~\ref{eqn:good_cnstr} have been applied, all angular observables
are still constructed out of products of spin-amplitudes in the
fixed-basis.

\subsection{Approximate discrete symmetries}
\label{sec:acc_sym}
\begin{figure}[!hbt]
\centering
\includegraphics[width=0.49\textwidth]{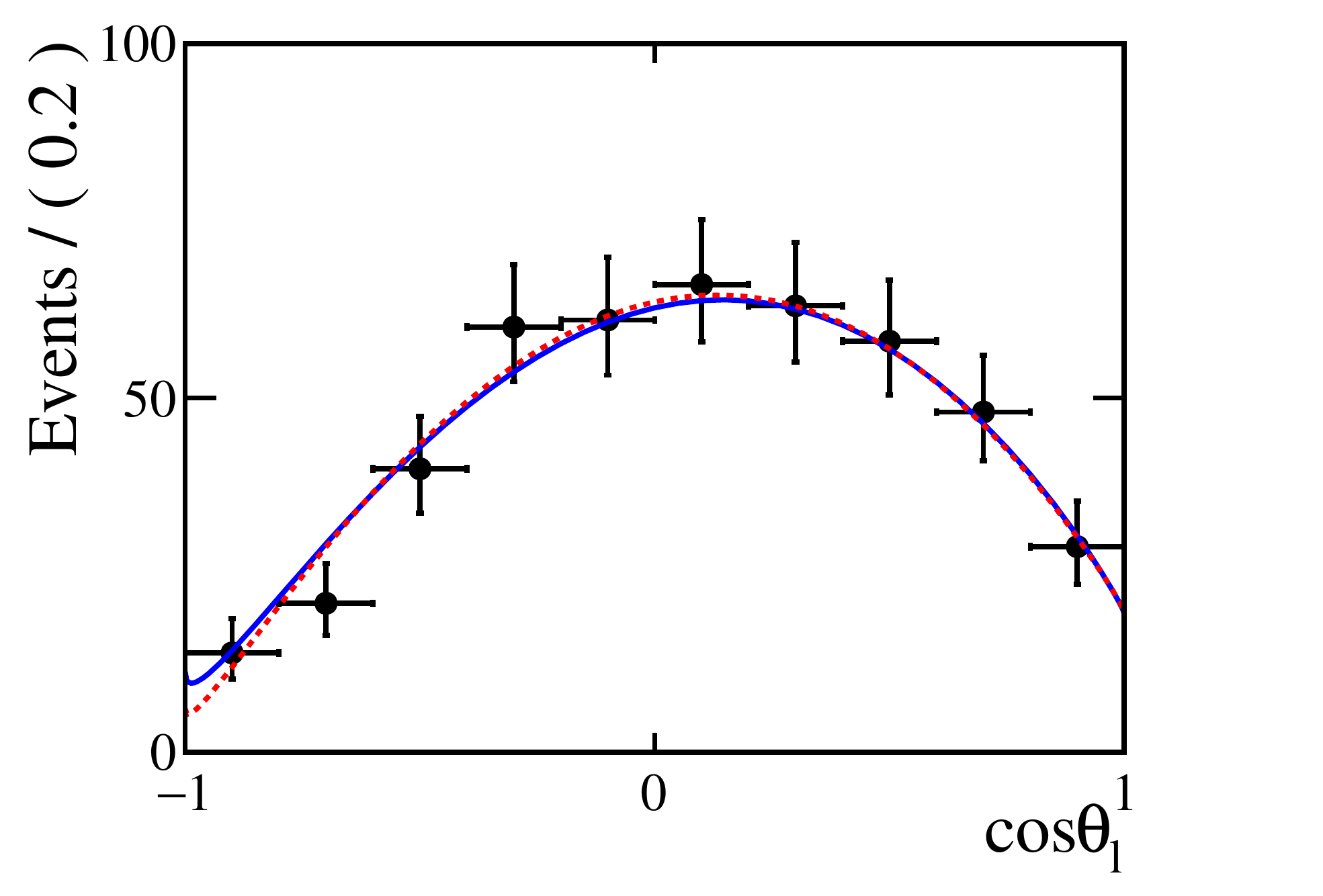}
\includegraphics[width=0.49\textwidth]{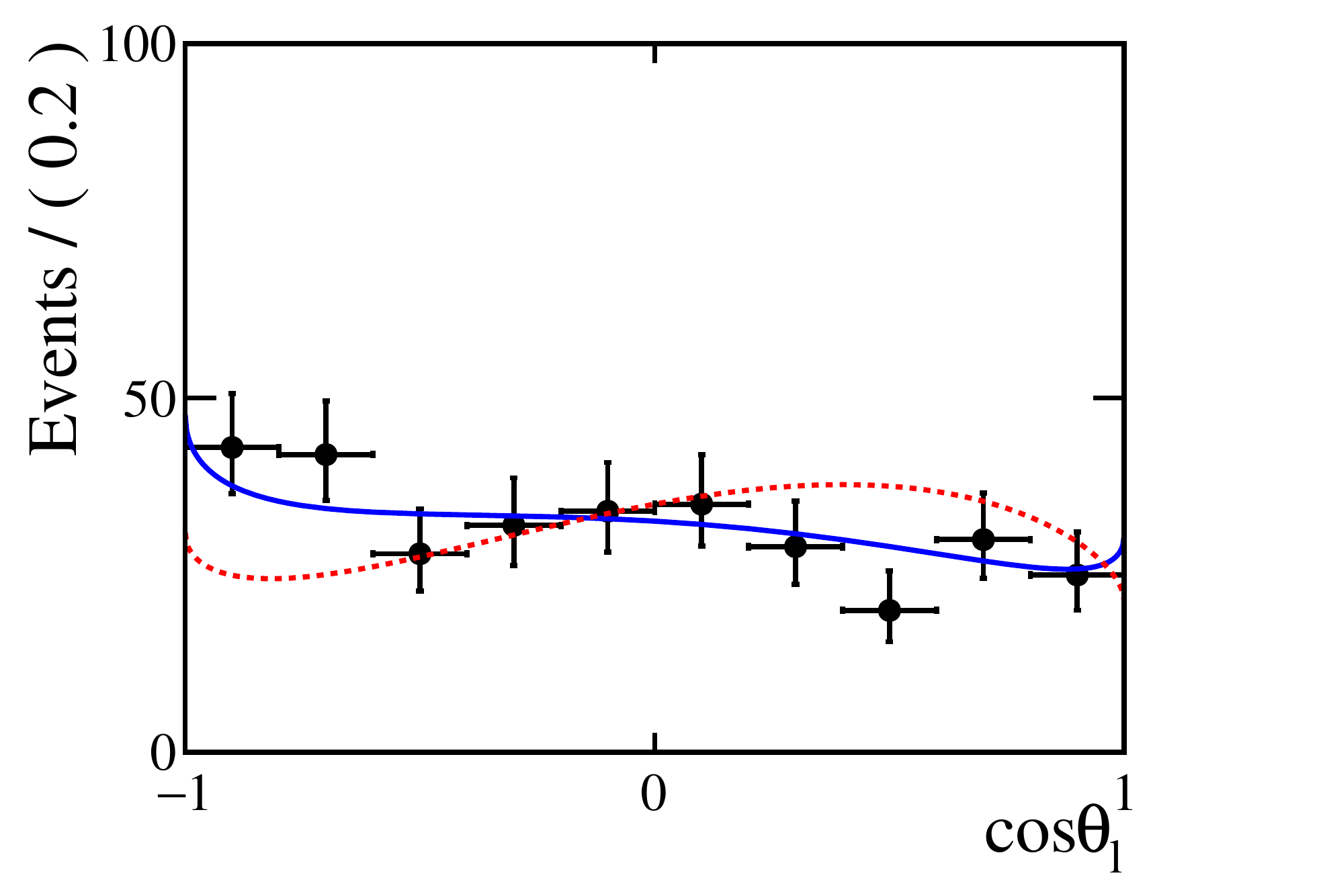}
\caption{ {\small Demonstration of the effect of the transformation of
    Eq.~(\ref{eqn:discr_symm}) in the SM (left) and in a model with
    large right-handed Wilson Coefficients (right). The amount of data
    corresponds to the expected number of \decay{\Bzb}{\Kstarzb\mumu}
    candidates in LHCb's Run-II dataset. The blue line denotes the
    model that the data is generated from. The red-dashed line denotes
    the model with the approximate symmetry transformation, as
    mentioned in the main text, applied. The angles $\phi$ and
    $\theta_{K}$ are required to satisfy $-\pi/4<\phi<\pi/2$ and
    $0<\cos\theta_{K}<1$.}  }\label{fig:acc_symm}
\end{figure}

The limited amount of signal candidates available in the experimental data,
can give rise to approximate symmetries under discrete transformations
of the amplitudes. The exact form of these approximate symmetries can
depend on the basis-fixing transformations discussed in
Sec.~\ref{sec:amplitudes_inf_sym}. Given the basis-fixing condition of
Eq.~(\ref{eqn:good_cnstr}), a clear example occurs in the transformed
basis with the SM amplitudes, where the lack of right-handed currents can give rise
to an approximate symmetry under the transformation
\begin{align}
\label{eqn:discr_symm}
A_{\parallel}^{L} &\rightarrow -A_{\perp}^{L}\nonumber\\
A_{\perp}^{L}&\rightarrow -\frac{A_{\parallel}^{L}}{2}
\end{align}
The effect of this accidental approximate symmetry can be demonstrated
by generating samples based on the SM and on a model with large
right-handed Wilson coefficients. Figure~\ref{fig:acc_symm}
shows the effect of the discrete transformation of
Eq.~\ref{eqn:discr_symm} on $\cos\theta_{\ell}$, both in the SM and
the model with large right-handed currents. The region where
$-\pi/4<\phi<\pi/2$ and $0<\cos\theta_{K}<1$ is considered, in order
to reduce possible cancellation of terms arising from the integration
of the angular distribution over $\phi$ and
$\cos\theta_{K}$. Figure~\ref{fig:acc_symm} shows that in the SM the
angular distribution in $\cos\theta_{\ell}$ is essentially
indistinguishable under the transformation; whereas, in the model with
right-handed currents, the angular distributions can be
distinguished. It is therefore clear that the above transformation is
an approximate discrete symmetry of the angular distribution only in
the SM and in other models with no right-handed currents.

An additional approximate discrete
symmetry exists under the transformation of the right handed
amplitudes in the transformed basis
\begin{align}
\label{eqn:discr_symm_rhanded}
A_{\parallel}^{R} &\leftrightarrow -A_{\perp}^{R}. 
\end{align}
\noindent As for the left-handed amplitudes, the transformation of
Eq.~(\ref{eqn:discr_symm_rhanded}) is an approximate discrete symmetry
of the angular distribution only in the SM and in other models with no
right-handed currents.

\subsection{Parameterised amplitudes} 
\label{sec:param_amps}
In order to determine the $q^2$ dependent \Kstarzb spin amplitudes, a
$q^2$ parametrisation of the amplitudes in the fixed-basis needs to be
employed. A three-parameter ansatz of the form
\begin{equation}
\label{eq:ansatz}
A=\alpha+\beta q^2+\gamma /q^2
\end{equation}
for both the real and imaginary components of each left- and right-
handed amplitudes is chosen. No attempt is made to interpret these
$\alpha$, $\beta$ and $\gamma$ coefficients in terms of short- or
long-distance parameters. 
The choice of this ansatz is justified by fitting the transformed spin
amplitudes, as provided by the \texttt{EOS}
program~\cite{Bobeth:2010wg}, in the SM and numerous other physics models, using
the parametrisation described above. Figure~\ref{fig:param_SM} shows
the result of the SM fit. Any bias coming from this choice of ansatz
will be much smaller than the statistical uncertainty of current and
any foreseeable-future experimental measurements.

The basis-fixing reduces the number of amplitude components that need
to be determined to eight per \Bz flavour.  Considering that each such
component is described by three parameters to account for the $q^2$
dependence, in total there are twenty-four amplitude parameters per
\Bz flavour that need to be determined. This parameter counting ignores
any S-wave amplitudes. Such amplitudes are discussed further in
Sec.~\ref{sec:swave_contribution}. Alternatively, the
model dependent assumption can be made, that the only weak phases present in the
amplitudes come from the CKM matrix elements. This assumption leads to
the $\Bz$ and $\Bzb$ amplitudes being identical, as the diagrams with
non-zero weak phases are Cabibbo suppressed. Accounting for the
experimental angular convention of the decay rate described in
Sec.~\ref{sec:diffdecayrate}, the decay distribution of both the $\Bz$
and $\Bzb$ decays can be described using a single set of amplitude
parameters. The approaches with separate and identical $\Bz$ and
$\Bzb$ amplitude parameters, are both discussed below.

\subsection{S-wave contribution}
\label{sec:swave_contribution}
Previous studies have discussed both the potential size as well as the
impact of the S-wave contribution in the angular analysis of
\decay{\Bz}{\Kstarz\mumu}
decays~\cite{ref:besirevic_swave,ref:hiller_shires_swave,ref:blake_egede_shires_swave}. In
particular, it has been shown that the number of signal
candidates expected in LHCb's Run-I dataset, ignoring the S-wave contribution
can have a significant effect on some angular observables.  It is
therefore critical that the $q^2$ dependent S-wave amplitude
components are also accounted for in the fit to the angular
distribution of \decay{\Bz}{\Kstarz\mumu} decays.

In this study, the S-wave amplitudes are included in the angular
distribution of the signal based on Eq.~\ref{eq:decrateswave} and are
treated as nuisance parameters in the fit. The $K^+\pi^-$ mass range
considered corresponds to 100~\mevcc around the $\Kstarz(892)$ pole
mass. The $m_{K\pi}$ dependence is accounted for by modifying each
$J_i$ term in Eq.~\ref{eq:decrateswave} by
\begin{equation}
J_{ij}=A_{i}A_{j}^{*}\to A_{i}A_{j}^{*}\int g_{i}(m_{K\pi})g_{j}^{*}(m_{K\pi})dm_{K\pi},
\end{equation}
where $g_{i}(m_{K\pi})$ represents the $m_{K\pi}$ line-shape of either
a P-wave or an S-wave amplitude. It is thus assumed that the
amplitudes do not explicitly depend on $m_{K\pi}$.

\begin{figure}[!htb]
\centering
\includegraphics[width=0.69\textwidth]{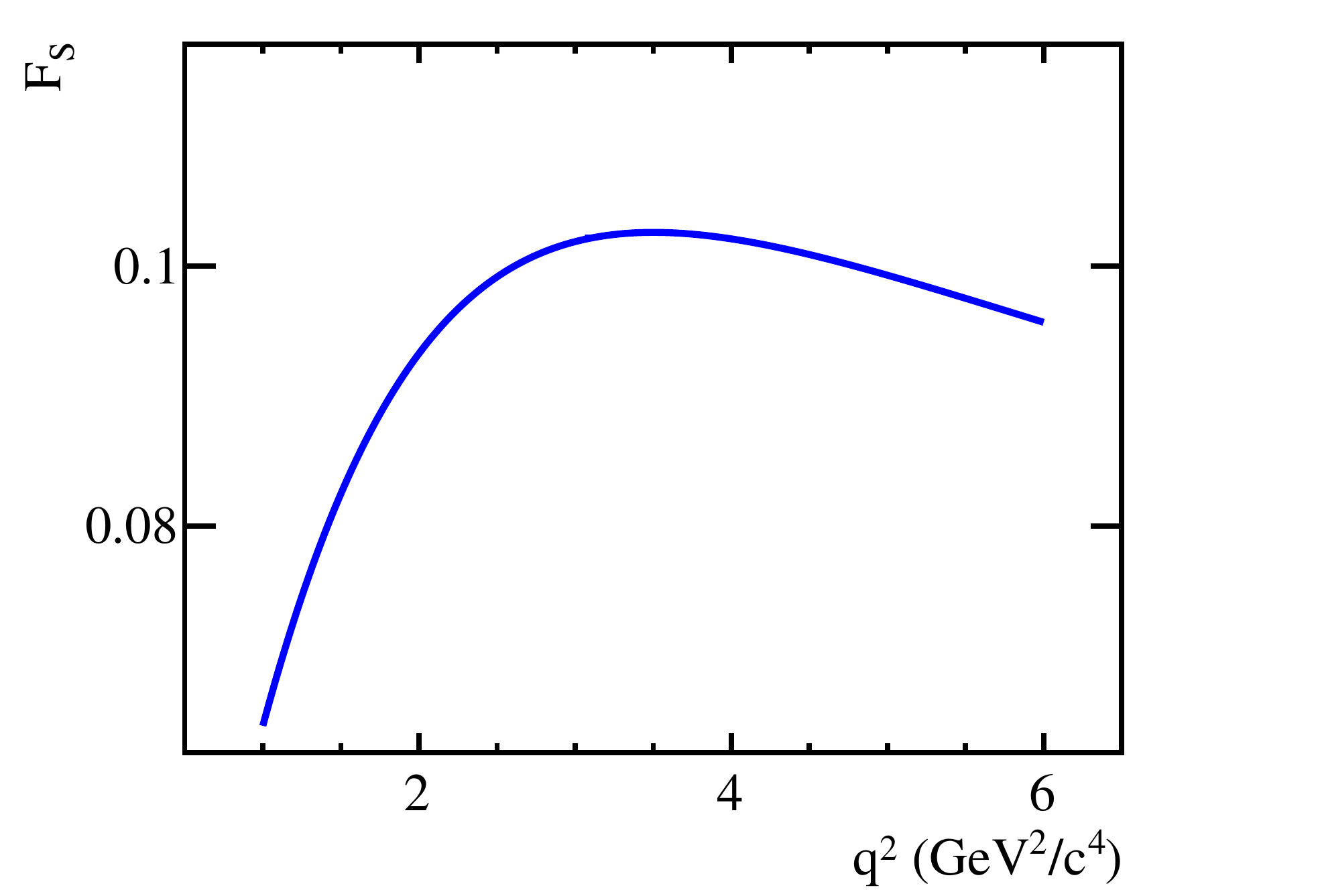}
\caption{ {\small Estimate of the S-wave fraction $F_S$ in the SM as
  a function of $q^2$ using the simplistic approach described in the text.}
\label{fig:fs_q2}
}
\end{figure}

The $q^2$ dependence of the S-wave amplitudes used in the generation
of the simulated events are calculated following
Ref.~\cite{ref:wang_swave}. For simplicity, only the $\kappa(600)$ is
considered to contribute to the S-wave in the $K^+\pi^-$ mass range
considered. This means that for this analysis only the line-shape of
the $\kappa$ is considered to contribute to the S-wave
$g_{i}(m_{K\pi})$. The line-shape of both the $\kappa(600)$ and the
$\Kstarz(892)$ are taken as relativistic Breit-Wigner distributions
with mass and width parameters as given in Ref.~\cite{PDG2012}.  The
form-factor of the $\kappa(600)$ is taken from
Ref.~\cite{ref:wang_kappa}.  The values of the corresponding
$\int g_{i}(m_{K\pi})g_{j}^{*}(m_{K\pi})dm_{K\pi}$ terms are shown in
Tab.~\ref{tab:mkpi_int}. In the SM, the resulting value of $F_S$ as a function of
$q^2$ is shown in Fig.~\ref{fig:fs_q2}. Using this
simplistic approach, the predicted value of $F_S$ is similar to the
values obtained using more sophisticated treatments, such as those of
Refs.~\cite{ref:besirevic_swave,ref:hiller_shires_swave}.

Given the size of current data samples, and the fact that the S-wave
fraction is expected to be small ($\mathcal{O}(10\%)$), the $q^2$
dependence of the S-wave amplitudes $A_{00}^{L,R}$, can be
approximated to be constant as a function of $q^2$ when performing
fits to the data. This is a good approximation since the $q^2$ shape
of the S-wave amplitudes is expected to be similar to that of
$A_{0}^{L,R}$~\cite{ref:wang_swave}, which is approximately constant
in the region $1<q^2<6$~\gevgevcccc.

\begin{table}
  \caption{Summary of the various integrals of the S- and P- wave
    line shapes that are used both in this study. The integral is
    performed in the $m_{K\pi}$ range $[796,996]~\gevcc$. }
\label{tab:mkpi_int}
\def\arraystretch{1.3}
\begin{center}
\begin{tabular}{cc}
\hline
Term & Value \\ 
\hline\hline
  $\int |g_{K^{*0}}|^2 dm_{K\pi}$ & $0.80$ \\
  $\int |g_{\kappa}|^2 dm_{K\pi}$ & $0.18$ \\
  $\int g_{\kappa}g_{K^{*0}}^{*}dm_{K\pi}$  &$0.22-0.23i$\\
\hline
\end{tabular}
\end{center}
\end{table}
\def\arraystretch{1.0}

\subsection{Determining the amplitudes}
\label{sec:extract_amplitudes}
The stability and sensitivity of a fit for the $q^2$ dependent
amplitudes is determined using simulated data with sample sizes
equivalent to those expected at LHCb during Run-I and Run-II of the
LHC. Estimates of the signal and background yields are taken from
Refs.~\cite{LHCb-PAPER-2013-019,LHCb-PAPER-2013-037} and scaled
linearly. Table~\ref{tab:yields}
summarises the signal and background yields used in the generation of
the simulated data samples. The angular distribution of the signal is
described using Eq.~(\ref{eq:decrateswave}), where the $q^2$ dependent
amplitudes are again calculated using the \texttt{EOS}
program~\cite{Bobeth:2010wg}, for both the SM and new physics models.
The angular distribution of the background is both generated and
described in the fit as a product of four one-dimensional functions,
each describing the dependence to the three helicity angles and $q^2$,
as shown in Eq.~(\ref{eq:totalmodel}),
\begin{equation}
  \label{eq:totalmodel}
  \begin{split}
    \frac{\deriv^{4}\Gamma[\mathrm{Bkg}]}{\deriv\cos\theta_{\ell}\,\deriv\cos\theta_{K}\,\deriv\phi\,\deriv
      q^{2}} & = f(\cos\theta_{\ell})\times g(\cos\theta_{K})\times
    h(\phi)\times l(q^2)~,
  \end{split}
\end{equation}
where $f,~g,~h,~l$ are first order polynomials. 

\begin{table}
  \caption{Summary of the signal and background yields used in
    the generation of the simulated data samples. Events are assumed
    to be equally distributed between $\Bz$ and $\Bzb$ candidates, both for signal
    and background.}
\label{tab:yields}
\begin{center}
\begin{tabular}{ccc}
\hline
            & \multicolumn{2}{c}{Yields} \\
Sample & Run-I & Run-II \\ 
\hline\hline
  Signal          & 600 & 2400 \\
  Background & 500 & 2000 \\
\hline
\end{tabular}
\end{center}
\end{table}

For each dataset, the amplitude coefficients of Eq.~\ref{eq:ansatz}
are determined using an extended maximum likelihood fit. The
probability distribution functions for the signal and background,
$P_{\mathrm{Sig(Bkg)}}$, are formed from the decay rate functions of
Eq.~\ref{eq:decrateswave} and Eq.~\ref{eq:totalmodel}, respectively.
The signal amplitudes, which make up the various $J_i$ factors of
Eq.~\ref{eq:decrateswave}, are written in terms of the three-parameter
ansatz of Eq.~\ref{eq:ansatz}. The likelihood function is thus

\begin{eqnarray}
  \label{eq:lhood}
  \displaystyle 
  -\log\mathcal{L} &=& \sum_{i}^{N_{\mathrm{Dat}}} -\log[N_{\mathrm{Sig}}(\alpha_j,\beta_j,\gamma_j)P_{\mathrm{Sig}}(\cos\theta_{\ell},\cos\theta_{K}, \phi, q^2)+\nonumber\\
  && \hspace{0.68in}N_{\mathrm{Bkg}}P_{\mathrm{Bkg}}(\cos\theta_{\ell},
  \cos\theta_{K},\phi, q^2)]+\nonumber \\\nonumber\\
  && -N_{\mathrm{Dat}}\log[N_{\mathrm{Sig}}(\alpha_j,\beta_j,\gamma_j)+N_{\mathrm{Bkg}}]+[N_{\mathrm{Sig}}(\alpha_j,\beta_j,\gamma_j)+N_{\mathrm{Bkg}}],
\end{eqnarray}
where $N_{\mathrm{Dat}}$ is the total number of events in the
dataset, $N_{\mathrm{Bkg}}$ is a parameter in the fit
that gives the number of background events, and
$N_{\mathrm{Sig}}(\alpha_j,\beta_j,\gamma_j)$ is the
number of signal events written in terms of the integrated signal
decay rate:
\begin{equation}
N_{\mathrm{Sig}}(\alpha_j,\beta_j,\gamma_j) = \frac{\bar{N}_{Dat}}{\Delta q^2}
\int_{-1}^{1}\int_{-1}^{1}\int_{-\pi}^{\pi}\int_{1\gevgevcccc}^{6\gevgevcccc}\frac{\deriv^{4}\Gamma[\mathrm{Sig}]}{\deriv\cos\theta_{\ell}\,\deriv\cos\theta_{K}\,\deriv\phi\,\deriv
  q^{2}}\,\deriv\cos\theta_{\ell}\,\deriv\cos\theta_{K}\,\deriv\phi\,\deriv
      q^{2},
\end{equation}
where $\bar{N}_{Dat}$ represents the average expected number of signal and
background events for a given LHCb data sample, and $\Delta q^2$
denotes the $q^2$ range in consideration, in this case
$5~\gevgevcccc$.

\section{Results}
\label{sec:amplitudes_results}
An ensemble of $10^4$ simulated data sets is generated containing
signal and background events as described in
Sec.~\ref{sec:extract_amplitudes}. A maximum likelihood fit is
performed to each of the data sets, to extract the $q^2$ dependent P-
and S-wave spin-amplitudes. Therefore at a given value of $q^2$,
$10^4$ determinations of each amplitude and thus of each angular
observable are performed.  The results of the fits for the $q^2$
dependent amplitudes are presented in two scenarios. Firstly for a
sample size equivalent to the full LHC Run-II data sample expected to
be collected by LHCb, where the amplitudes of the $\Bz$ and $\Bzb$ are
both extracted without any model dependent assumptions (Scenario-II);
and secondly, for a sample size equivalent to the data sample
collected by LHCb during Run-I of the LHC, where it is assumed that
all weak phases of the amplitudes can be safely neglected
(Scenario-I).  In the latter case, the model dependent choice allows
the sensitivity to the \CP-averaged observables, $S_i$, to be
maximised, with the smaller sample that will be available from Run-I
of the LHC.

\begin{figure}[!hbt]
\centering
\includegraphics[width=0.98\textwidth]{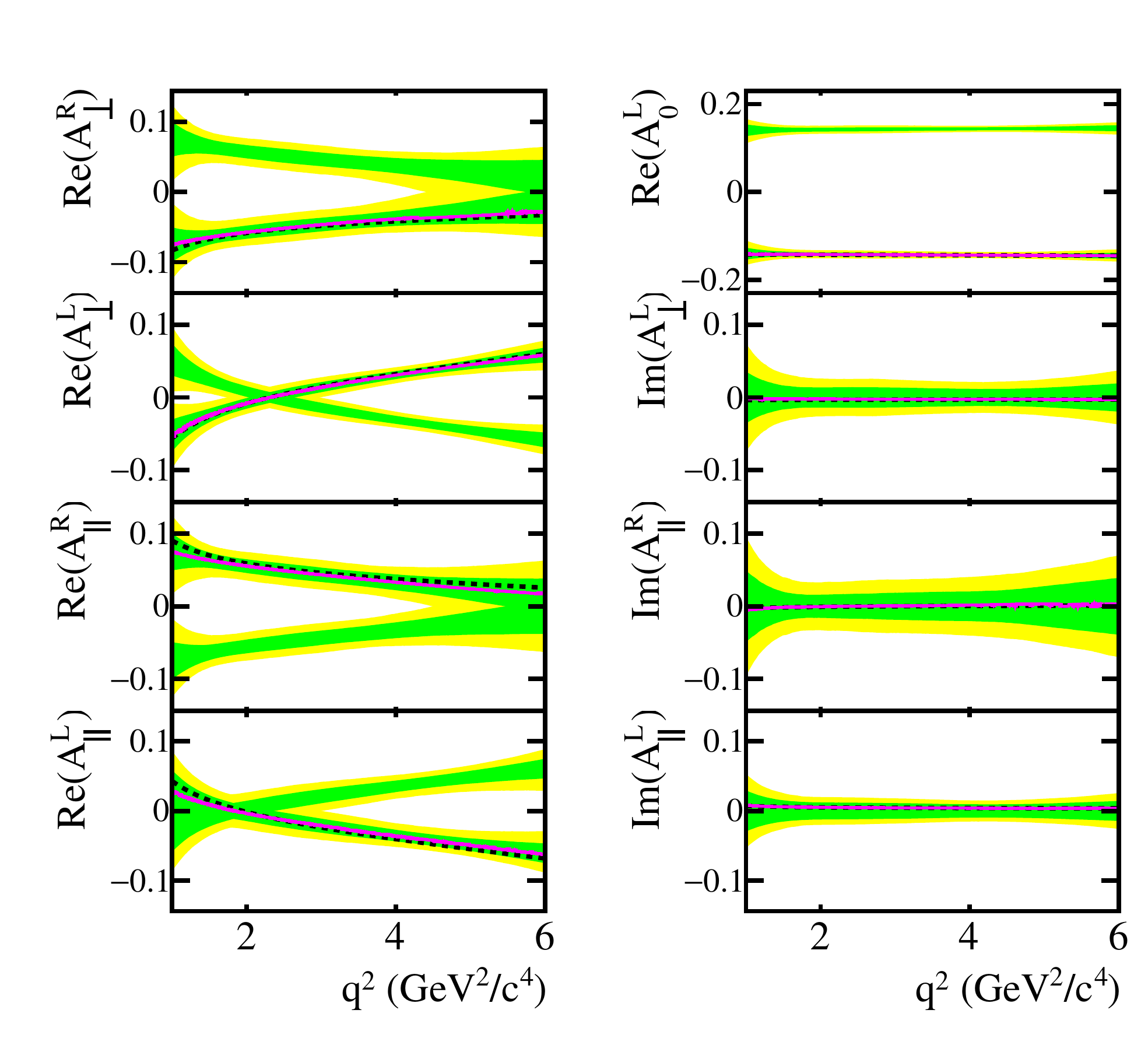}
\caption{ {\small Distributions of the \Bzb amplitudes as a
    function of $q^2$ resulting from $10^4$ fits to generated signal
    and background simulated data under Scenario-II. The green and yellow
    bands correspond to the 68\% and 95\% confidence intervals
    respectively.  The dotted line denotes the SM prediction as
    given by the \texttt{EOS} program~\cite{Bobeth:2010wg}. The magenta solid line denotes the most likely
    value resulting from the ensemble of fits. A discrete symmetry of
    reflections about zero is observed due to the fact that the $J_i$
    terms are bilinear coefficients of the amplitudes.  }
\label{fig:amps_param_SM}
}
\end{figure}

The resulting $q^2$ dependent P-wave \Bzb amplitudes, obtained from
fits to an ensemble of simulated data under Scenario-II, are shown in
Fig.~\ref{fig:amps_param_SM}.  At a given point in $q^2$, the 68\% and
95\% confidence intervals can be computed.  Connecting these points at
different $q^2$ values gives the statistical uncertainty on the
amplitudes as a function of $q^2$. A clear degeneracy is observed
under reflections about the x-axis. This effect is a consequence of
the discrete symmetry $A_i\to -A_i$, as discussed in
Sec.~\ref{sec:disc_sym}. Given the observable quantities are bilinear
combinations of the amplitudes, there is no corresponding degeneracy
in any observable.

\begin{figure}[!hbt]
  \centering
  \includegraphics[width=0.98\textwidth]{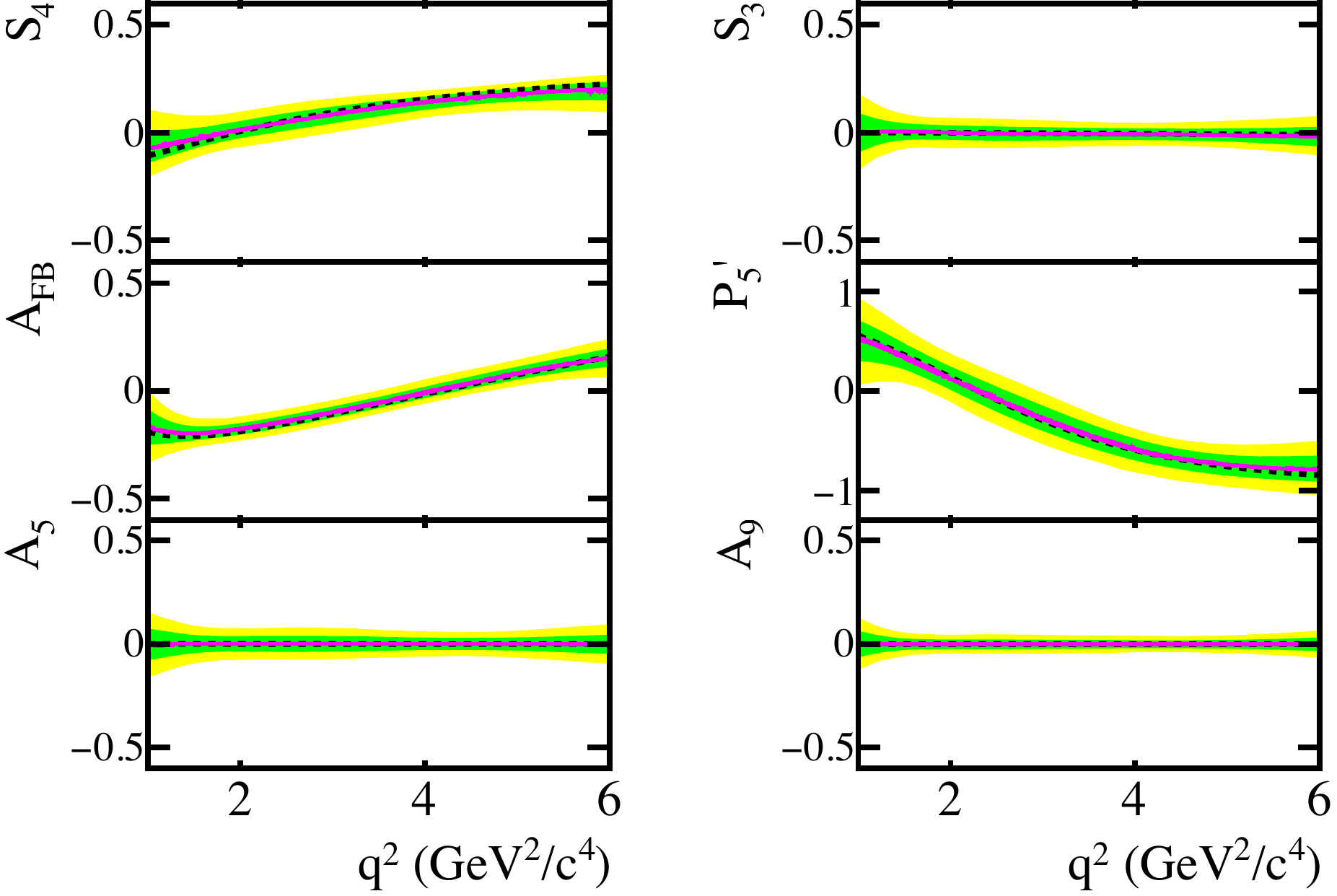}
  \caption{ {\small Subset of observables constructed out of
      the helicity-amplitudes, resulting from fits to ensembles of
      simulated data, under Scenario-II. The meaning of the various
      bands and curves is given in Fig~\ref{fig:amps_param_SM}.}
    \label{fig:obs_param_scen_II_SM}
  }
\end{figure}

\begin{figure}[!hbt]
\centering
  \includegraphics[width=0.98\textwidth]{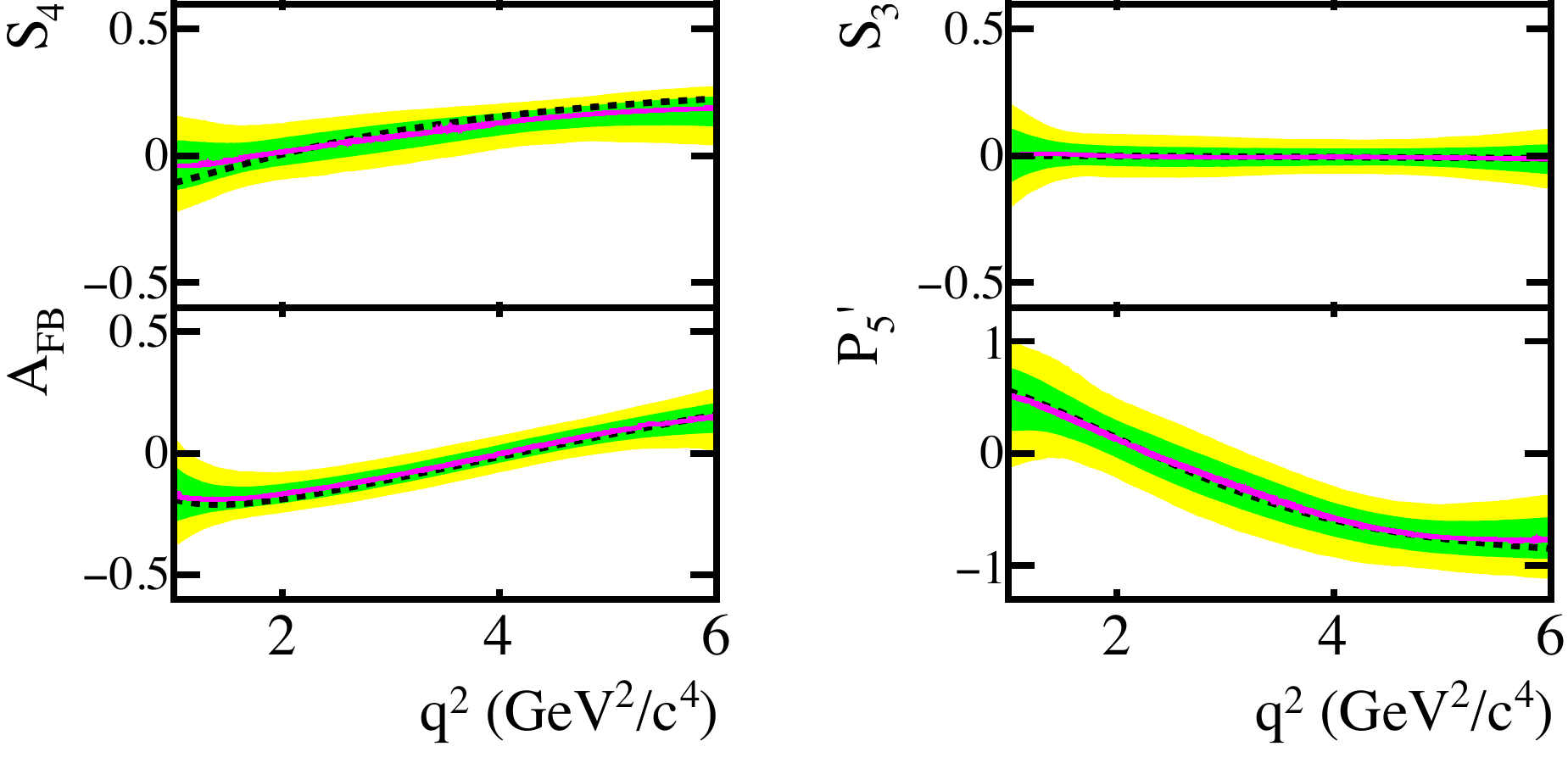}
  \caption{ {\small Subset of observables constructed out of
      the helicity-amplitudes, resulting from fits to ensembles of
      simulated data, under Scenario-I. The meaning of the various
      bands and curves is given in Fig~\ref{fig:amps_param_SM}.}
    \label{fig:obs_param_scen_I_SM}
  }
\end{figure}

The complete set of \CP-symmetric and \CP-asymmetric
observables, defined in Sec.~\ref{sec:diffdecayrate}, can be
constructed out the \Bzb and \Bz amplitudes. A subset of these
observables are shown in Figs.~\ref{fig:obs_param_scen_II_SM}
and~\ref{fig:obs_param_scen_I_SM}, for Scenario-II and Scenario-I
respectively. The 68\% and 95\% bands are given as in
Fig.~\ref{fig:amps_param_SM}.

In a minor part of the $q^2$ range, a bias at the level of $0.5\sigma$
is apparent in some observables, most notably $S_4$. The $S_4$
observable is sensitive to the interference between the
$A_{\parallel}^{L}$ and $A_{0}^{L}$ amplitudes which in the
fixed-basis is given by $\mathrm{Re}(A_{\parallel}^{L}A_{0}^{L})$.
This bias arises from the approximate discrete symmetry discussed in
Sec.~\ref{sec:acc_sym}. This accidental symmetry is more prone to
occur in models where the right-handed Wilson Coefficients are zero,
such as the SM. As this is an effect of the angular distribution, this
type of bias would have to be taken into account for any fitting
method employed, whether binned or unbinned in $q^2$.

\subsection{Uncertainty estimation}
The probability distribution function of the signal decay remains
positive definite for all values of the amplitude coefficients. This
fact, coupled with the expected LHCb sample sizes for Scenarios-I and
II, means that the likelihood surface, constructed out of the data and
the probability distributions of the signal and the background, should
be a good estimator of the statistical uncertainty of the amplitude
coefficients. 

Table~\ref{tab:pull_scen_B} summarises the pull mean and width of the
amplitudes in the fixed-basis for Scenario-I at a point in $q^2$.  In
order to discern any statistical bias from the bias originating from
the approximate symmetries discussed in Sec.~\ref{sec:acc_sym}, a
point in $q^2$ is chosen such that minimises the bias from the approximate
symmetries.

\begin{figure}[!htb]
\centering
\includegraphics[width=0.49\textwidth]{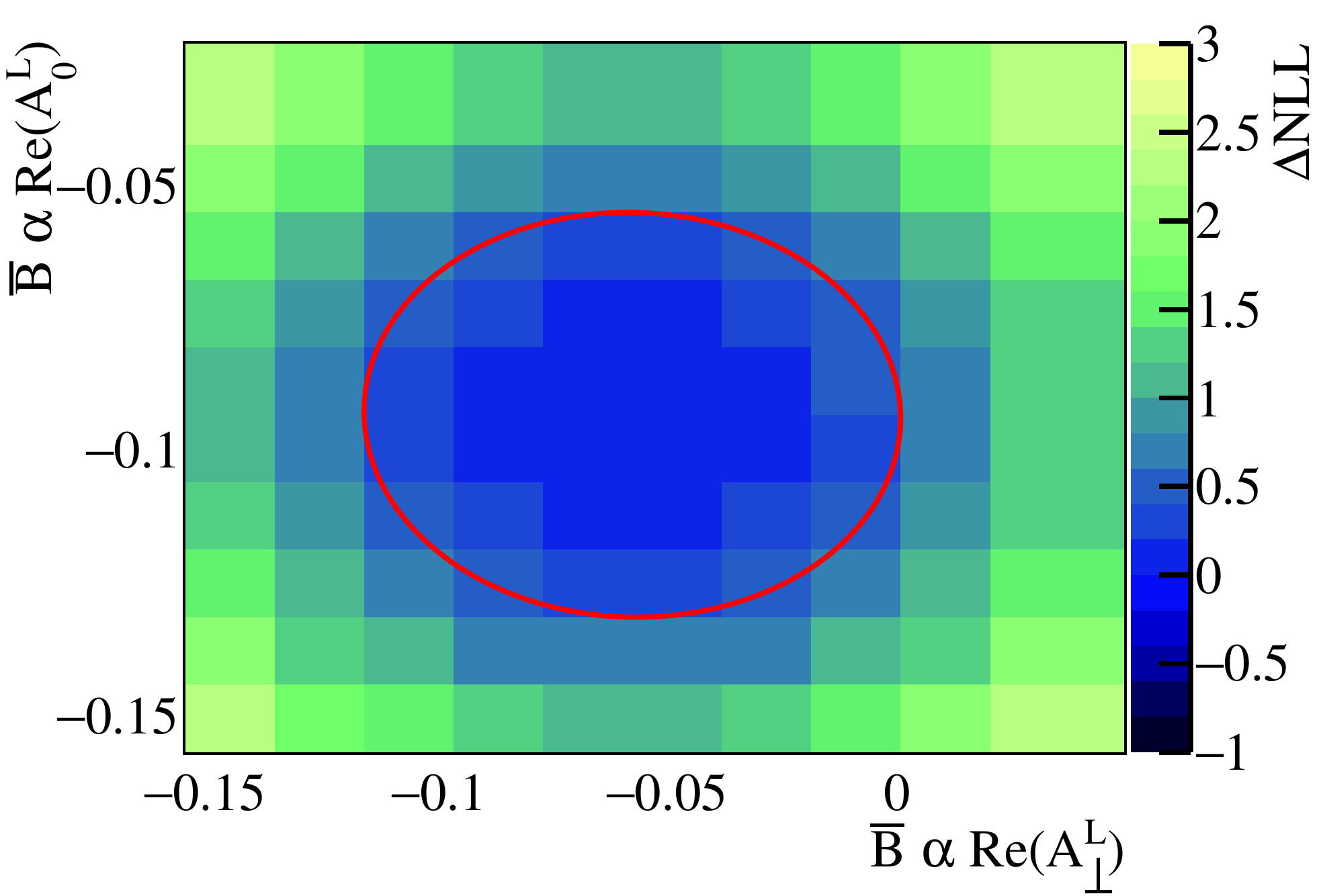}
\includegraphics[width=0.49\textwidth]{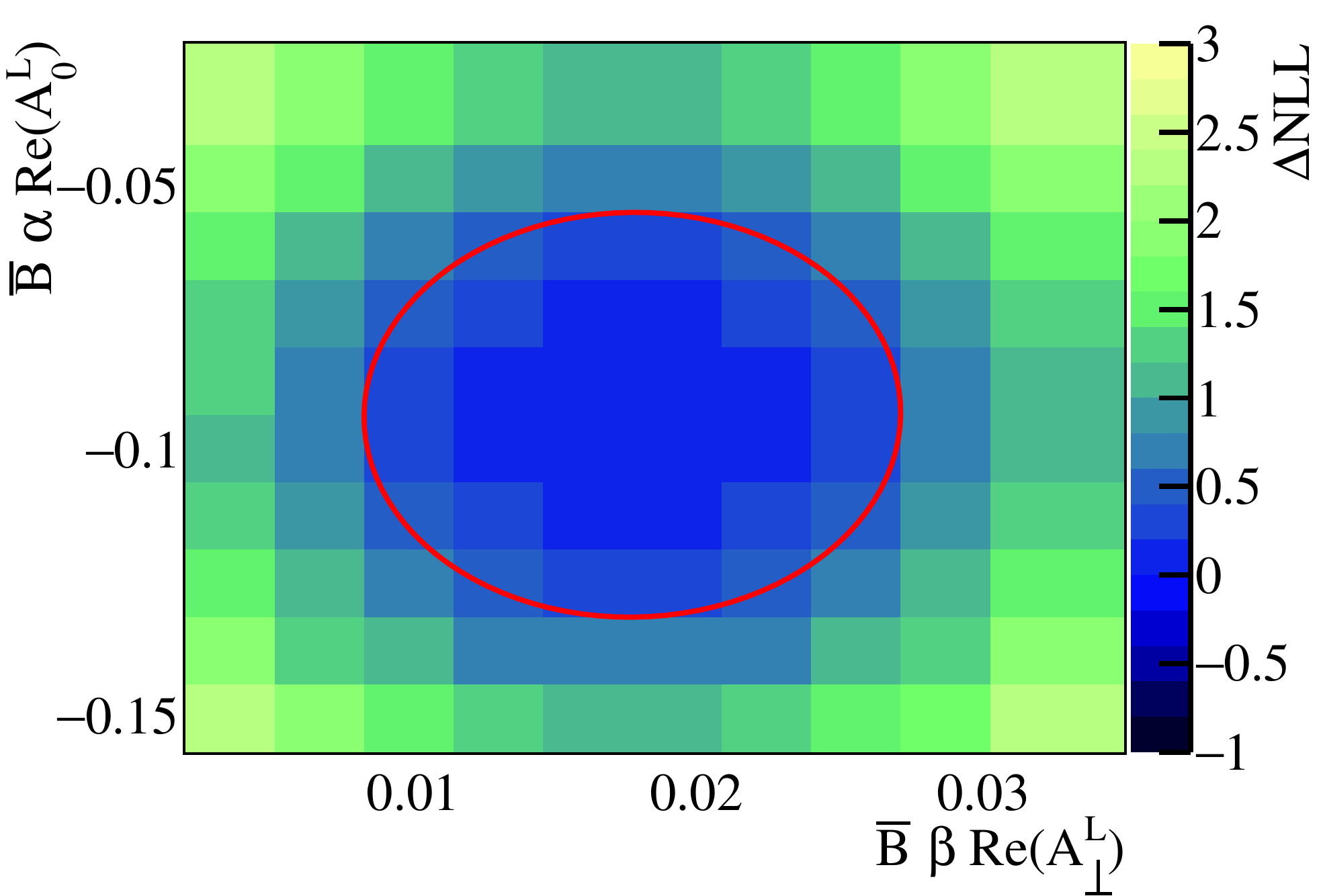}\\
\includegraphics[width=0.49\textwidth]{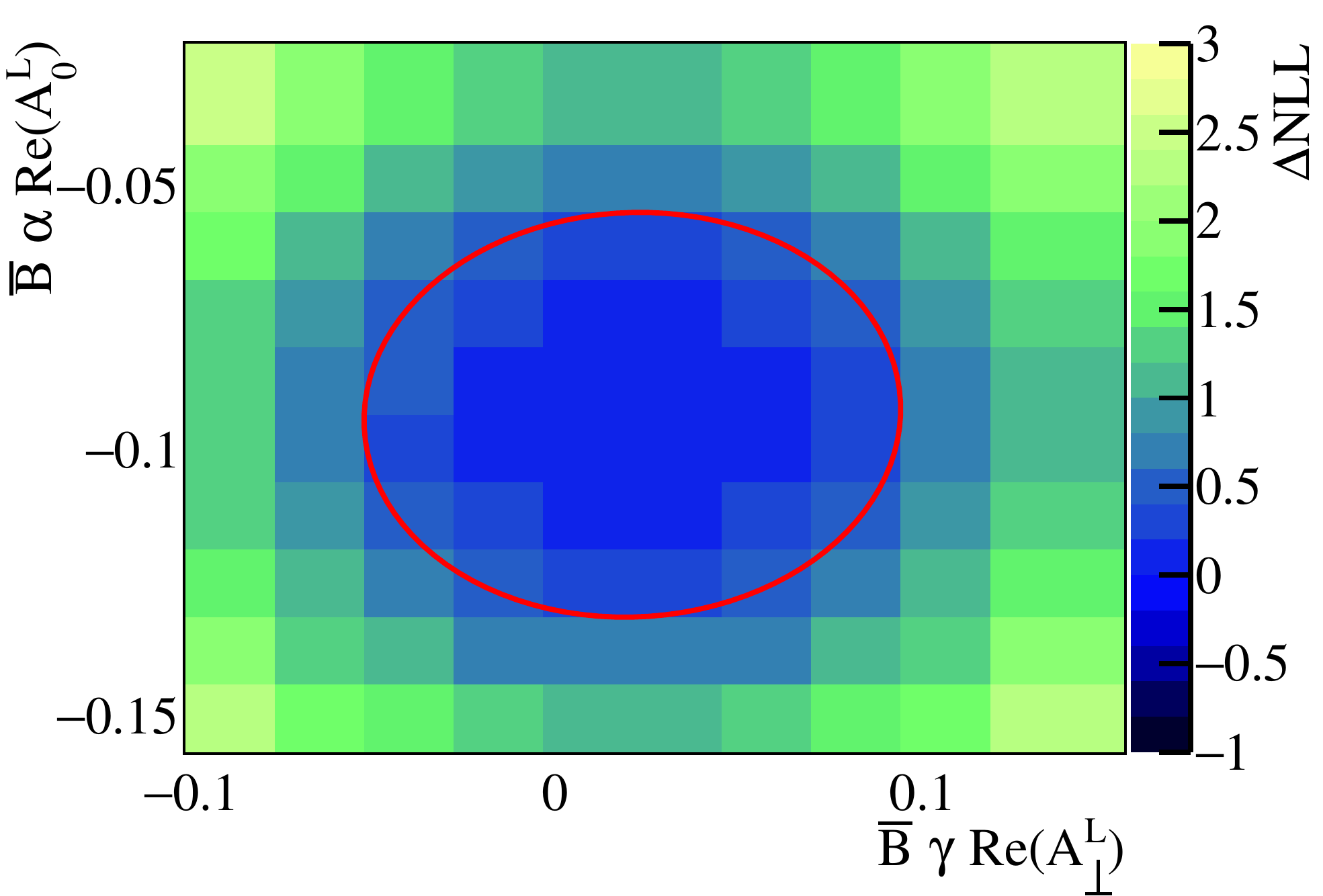}
\includegraphics[width=0.49\textwidth]{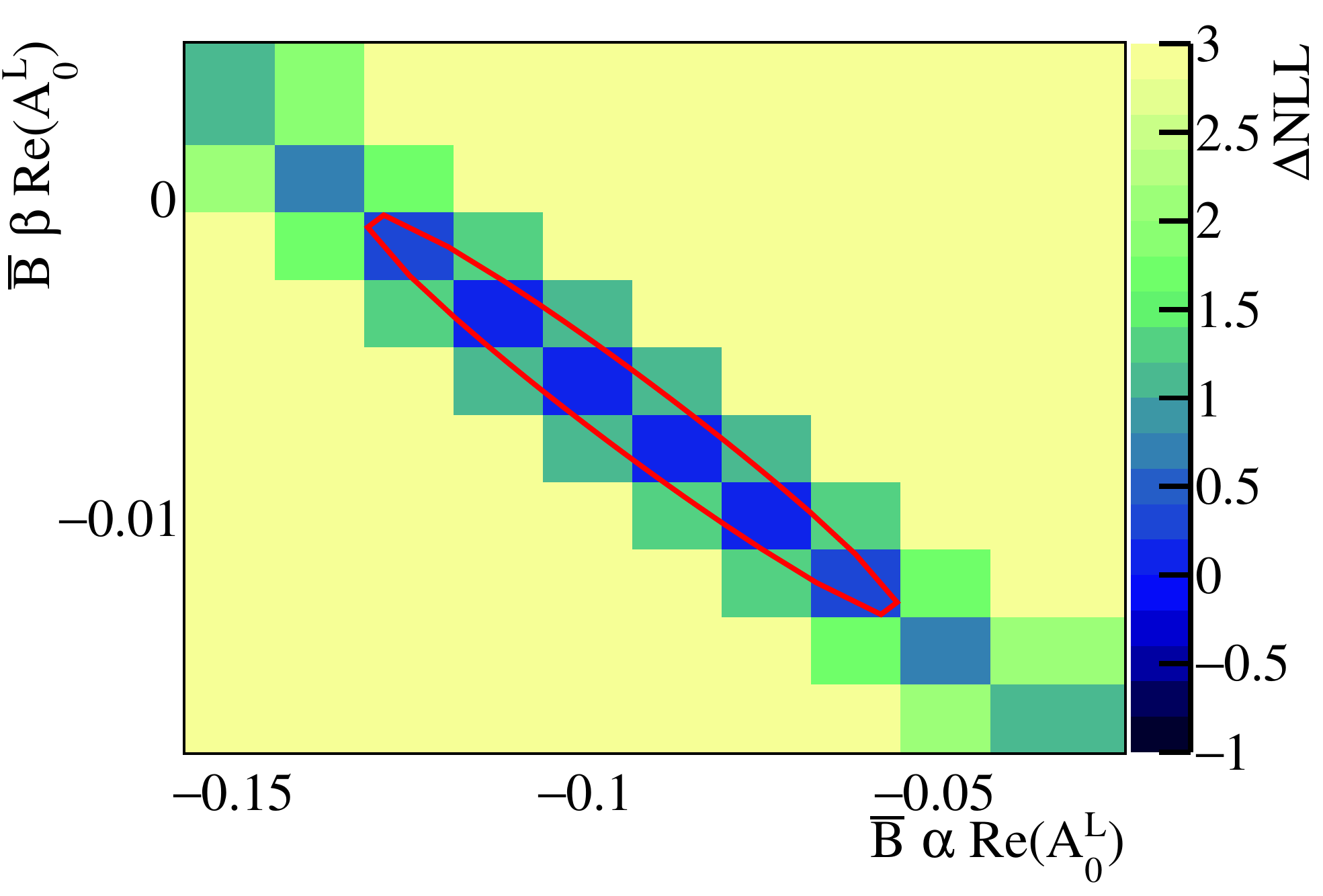}\\
\includegraphics[width=0.49\textwidth]{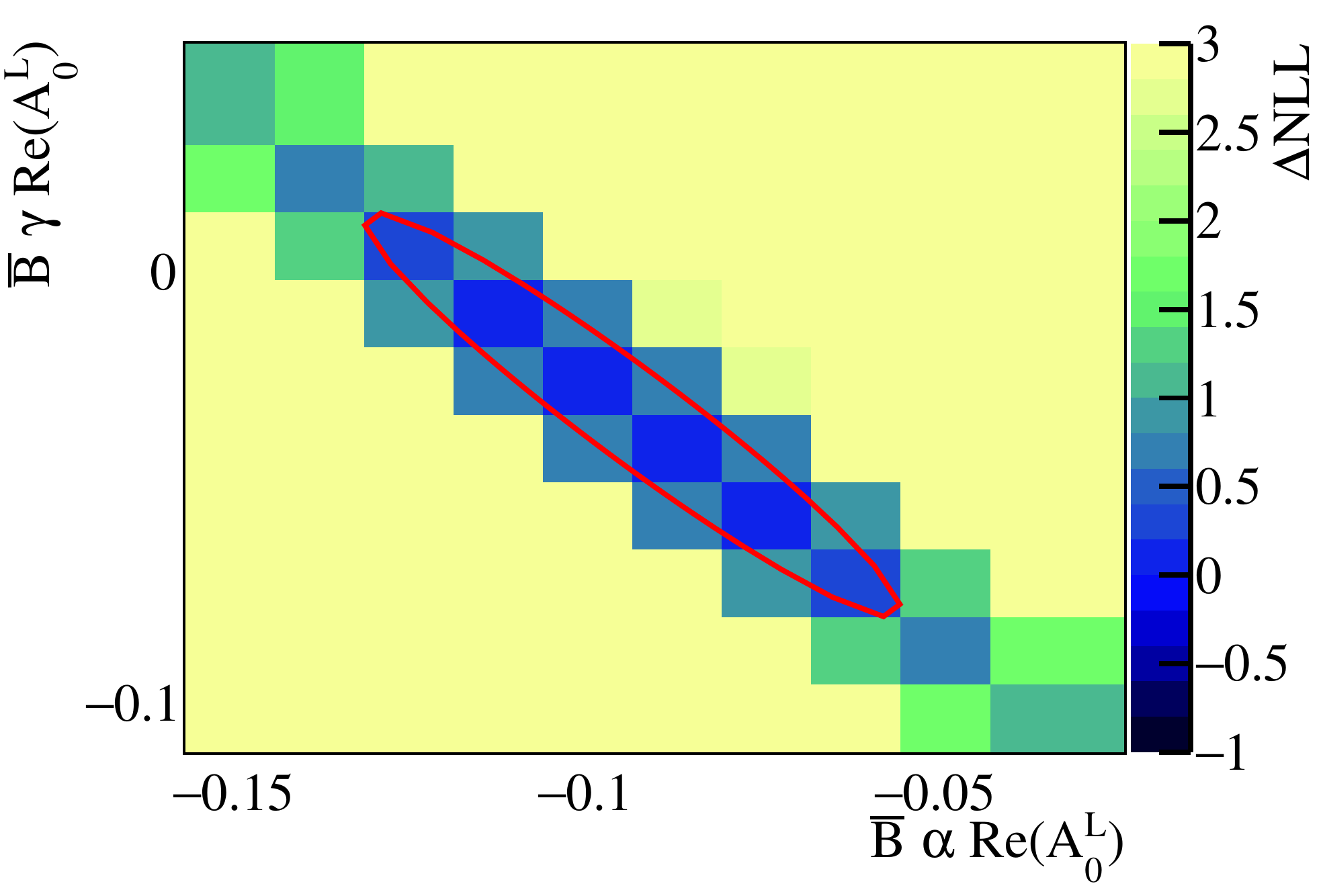}
\caption{ {\small Subset of two dimensional profile likelihood
    distributions from a fit to a single simulated dataset in
    Scenario-I. The red ellipses denote the error and the correlation
    obtained from the error matrix of the fit.}
\label{fig:prof_2d_scen_B}
}
\end{figure}

The pull means and widths are largely consistent with zero and unity,
respectively, indicating that the likelihood is a good estimator of
the uncertainty of the amplitudes.  The residual bias in
$A_{\parallel}^{R}$ arises from the additional approximate symmetry
between $A_{\parallel}^{R}$ and $A_{\perp}^{R}$ discussed in
Sec.~\ref{sec:acc_sym}. Figure~\ref{fig:amps_param_SM} shows that
there is no point in $q^2$ where the bias from the approximate symmetry can be
removed for all the amplitudes simultaneously.
\def\arraystretch{1.3}
\begin{table}
  \caption{\small Means and widths of the pull distributions of the P-wave
    amplitudes at $q^2=2.4$~\gevgevcccc, obtained from fits to ensembles of simulated data
    samples in Scenario-I. The pull is defined as (Fit-SM)/$\sigma_{\mathrm{Meas.}}$
    where $\sigma_{\mathrm{Meas.}}$ is the error on the measured
    quantity obtained using the error matrix of the fit. 
    The deviation of the pull mean of $\mathrm{Re}(A_{\parallel}^{R})$ from zero arises due to the 
    residual bias from the approximate symmetry of the angular distribution.}
\label{tab:pull_scen_B}
\begin{center}
\begin{tabular}{ccc}
\hline
Parameter                            & Pull mean               & Pull width\\ 
\hline\hline
$\mathrm{Re}(A_{0}^{L}) \Bzb$          & $-0.03 \pm 0.02$  & $0.97 \pm 0.03$\\
$\mathrm{Re}(A_{\parallel}^{L}) \Bzb$ & $\phantom{-}0.00 \pm 0.02$    & $1.01 \pm 0.03$\\
$\mathrm{Im}(A_{\parallel}^{L}) \Bzb$ & $\phantom{-}0.01 \pm 0.02$    & $1.02 \pm 0.03$\\
$\mathrm{Re}(A_{\parallel}^{R}) \Bzb$ & $\phantom{-}0.22 \pm 0.02$    & $0.90 \pm 0.03$\\
$\mathrm{Im}(A_{\parallel}^{R}) \Bzb$ & $-0.02 \pm 0.02$  & $0.94 \pm 0.03$\\
$\mathrm{Re}(A_{\perp}^{L}) \Bzb$     & $\phantom{-}0.02 \pm 0.02$    & $0.97 \pm 0.03$\\
$\mathrm{Im}(A_{\perp}^{L}) \Bzb$     & $-0.04 \pm 0.02$  & $0.95 \pm 0.03$\\
$\mathrm{Re}(A_{\perp}^{R}) \Bzb$     & $-0.05 \pm 0.02$  & $0.94 \pm 0.03$\\
\hline
\end{tabular}
\end{center}
\end{table}
\def\arraystretch{1.0} 
A subset of two dimensional profile-likelihood
distributions is shown in Fig.~\ref{fig:prof_2d_scen_B}. The profile
likelihood is obtained by scanning over the two parameters in
question, and minimising the likelihood over the rest of the parameters at
each point. These profile likelihoods can be used to determine the
confidence regions of the amplitude coefficients and therefore of the
observables. The error matrix of the fit is a good approximation to
the likelihood surface at the level of around $15\%$.

For a given dataset, a prediction of the spin-amplitudes as a function
of $q^2$ can be obtained for a particular value of the Wilson
Coefficients and $B\to K^{*}$ form factors. In turn, each amplitude
can be expressed in terms of the three parameter ansatz of
Eq.~(\ref{eq:ansatz}), by transforming to the fixed-basis, using the
procedure detailed in Appendix~\ref{app:a}. By fitting each transformed
amplitude with the ansatz of Eq.~(\ref{eq:ansatz}), the coefficients
$\alpha$, $\beta$ and $\gamma$ can be determined. The best fit point
and the error matrix of the fit to the data, can then be used to place
constraints on the Wilson Coefficients using a procedure like that
employed in Refs.~\cite{Descotes-Genon:2013wba,Altmannshofer:2013foa,
  Beaujean:2013soa,Mahmoudi:2014mja,Jager:2014rwa}.

\section{Sensitivity to new physics}
\label{sec:amplitudes_comparison}
The expected sensitivity to the effects of new physics (NP),
neglecting theory uncertainties, is estimated by generating a large
number of simulated data samples according to the SM (null dataset)
and a NP model where $\delta C_9=-1.5$ (test dataset). The choice of
the NP model is motivated by recent results from global fits to
$b\to s\ell\ell$ and $b\to s\gamma$
measurements~\cite{Descotes-Genon:2013wba,Altmannshofer:2013foa,Beaujean:2013soa,Hurth:2013ssa,Jager:2012uw,
  Jager:2014rwa,Descotes-Genon:2014uoa,Altmannshofer:2014cfa,Crivellin:2015mga,
  Gauld:2013qja,Altmannshofer:2014rta,Mahmoudi:2014mja, Datta:2013kja}
which are dominated by LHCb's anomalous results in the angular
distribution of \decay{\Bzb}{\Kstarzb\mumu}
decays~\cite{LHCb-PAPER-2013-037,LHCb-CONF-2015-002}. The \texttt{EOS}
program~\cite{Bobeth:2010wg} is used to generate simulated data from
these two models using their central value predictions. Two fits are
then performed to an ensemble of simulated datasets generated with the
SM and the NP model. In the first fit, the amplitude parameters are
fixed to their SM values (null hypothesis). In the second fit, the
amplitude parameters are fixed to the values given by the model with
$\delta C_9=-1.5$ (test hypothesis).  The background components and
yields are treated as nuisance parameters and are left floating in
each fit.  The S-wave contribution in the $K\pi$ system of
\decay{\Bzb}{\Kstarzb\mumu} decays is less understood theoretically than
the dominant P-wave part. In order to correctly model the S-wave
component of the $K\pi$ system, experimental input is
required. Therefore, for these sensitivity studies, the S-wave
amplitudes are treated as nuisance parameters in the fit.

\begin{figure}[!hbt]
\centering
\includegraphics[width=0.69\textwidth]{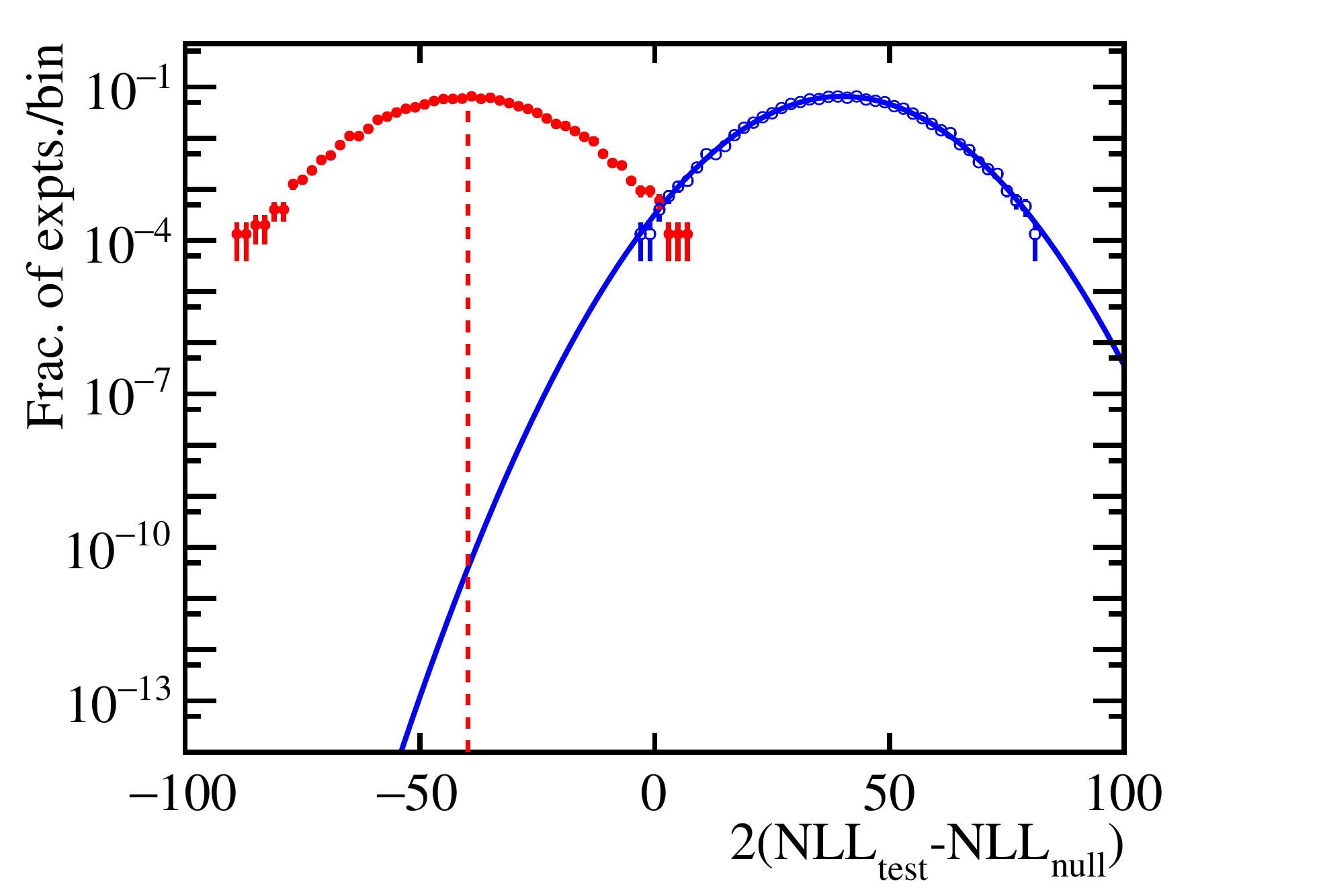}
\caption{ {\small Distribution of the test statistic $Q^{NP,SM}$, for fits to the
    NP (red full points) and SM (blue open
    points) simulated data samples. The S-wave
    components are treated as nuisance parameters and are therefore
    floating in the fit. The dashed red line denotes the median of the test
    statistic from fits to the NP sample. The solid blue curve
    is a fit of a Gaussian distribution to the distribution of the
    test statistic from fits to the SM sample.}
\label{fig:hypo_sens}
}
\end{figure}

The test statistics are defined as
\begin{align}
Q^{SM} & =2(\mathrm{NLL}^{SM}_{{test}}-\mathrm{NLL}^{SM}_{{null}})\nonumber\\
Q^{NP} & =2(\mathrm{NLL}^{NP}_{{test}}-\mathrm{NLL}^{NP}_{{null}}),
\end{align}
where $\mathrm{NLL}^{SM,NP}_{{null,test}}$ corresponds to the negative
log likelihood value of the null or test hypothesis on a SM or NP
simulated dataset. The expected sensitivity to a model with
$\delta C_9=-1.5$ is then estimated by counting the fraction of the
toy simulations with a value of $Q^{SM}\leq \bar{Q}^{NP}$, where
$\bar{Q}^{NP}$ is the median of the $Q^{NP}$
distribution. Figure~\ref{fig:hypo_sens} shows the distribution of the
test statistic for both the SM and NP simulated data samples in fits
to $\Bz$ and $\Bzb$ candidates separately, using a sample size
equivalent to that expected at LHCb during Run-I of the LHC. The
probability for the SM sample to fluctuate such that it gives a test
statistic as low or lower than the median of the NP sample (i.e
$Q^{SM}\leq \bar{Q}^{NP}$) corresponds to a significance of
$6.5\sigma$.  This significance is obtained for an idealised model of
the experimental data and does not account for the theoretical
uncertainties associated with the translation from Wilson coefficients
to amplitudes.

For comparison with methods used previously, the same
procedure for estimating the expected significance can be performed
for fits directly to the \CP-averaged observables, $S_i$ in bins of
$q^2$.  For these fits, three $q^2$ bins between $1<q^2<6$~\gevgevcccc
are chosen as $(1,2.7)$, $(2.7,4.3)$, $(4.3,6)$~\gevgevcccc. In this
binned approach a combined significance of $5.0\sigma$ is obtained.
Therefore, fitting for the $q^2$ dependent \Kstarz spin-amplitudes,
separately for the $\Bz$ and the $\Bzb$, results in a 30\% improvement
in the expected sensitivity. Equivalently, to get the same sensitivity
as the amplitude method, a binned fit to the \CP-averaged observables
would require an additional 70\% of integrated luminosity.

The inclusion of the $K\pi$ system in an S-wave configuration
introduces six additional observables, as shown in
Eq.~(\ref{eq:decrateswave}), in contrast to four additional S-wave
amplitude components.  Approximately half of the increase in
sensitivity detailed above, can be attributed to the reduced number of
S-wave-related nuisance parameters present in the amplitude fits, and
the other half owes to the intrinsic sensitivity of the method.

\section{Conclusions}
\label{sec:Conclusions}
In summary, a method of analysing \decay{\Bzb}{\Kstarzb\mumu} decays
is presented that allows the determination of all of the 
\Kstarz amplitudes as a parametric function of $q^2$. The method is
applicable with the data sample that is already available at the LHCb
experiment and works in the region where the amplitudes can be
described by a simple functional form, $1<q^2<6$~\gevgevcccc. The bias
coming from the choice of $q^2$ parameterisation is much smaller than
the statistical uncertainty of current and any forseeable-future
experimental measurements.

The method overcomes several shortcomings of previous methods: As all
the \Kstarzb helicity amplitudes are determined from a single fit to
data, the full correlations between experimentally determined
quantities can be obtained, improving the sensitivity to new physics;
fitting for the amplitudes enables all of the symmetries of the
angular distribution for a $K\pi$ system in a P- and S-wave
state to be accounted for, giving increased experimental precision
compared to previous approaches which retain redundant degrees of
freedom; the method avoids the integration of theoretical predictions
over experimental $q^2$ bins, enabling the full exploitation of the
cancellation of the form factors at leading order and therefore
further improving the sensitivity to new physics.

The continuous symmetry transformations that can be applied to the
amplitudes enable a modified basis to be specified, in which a number
of amplitude components are set to zero. The basis considered by
previous studies~\cite{ref:egede_matias_reece} gave a discontinuous
shape in $q^2$, rendering the region $q^2<2.5$~\gevgevcccc
unusable. Here, a new choice of basis is presented, which leaves the
amplitudes smoothly varying in the entire range $1<q^2<6$~\gevgevcccc
for the SM and for a wide range of new physics models. The present
analysis also highlights new approximate discrete symmetries that are
manifest for data samples on the order of that collected at LHCb
during Run-I. These approximate symmetries are more prone to occur in
models like the SM, where there are no right-handed currents.

In order to illustrate the options that are available to fit a given
dataset, the sensitivity of the method is presented in two different
scenarios. For a sample equivalent to the LHCb Run-I dataset, a
model-dependent assumption is made that the only weak phases present
in the amplitudes come from the CKM matrix elements. Given diagrams
with non-zero weak phases are Cabibbo suppressed, this assumption
results in $\Bz$ and $\Bzb$ amplitudes which are identical. The
\CP-averaged observables can then be determined with greater precision
than would be possible if instead both the \CP-averaged and
\CP-asymmetric observables were determined. For a sample equivalent to
the LHCb Run-II dataset, the results are determined without this
assumption and the sensitivity to both \CP-averaged and \CP-asymmetric
observables is presented.

The recent anomalous result in the angular distribution of
\decay{\Bzb}{\Kstarzb\mumu} motivates the consideration of a new
physics scenario with $\delta C_9=-1.5$~\cite{LHCb-PAPER-2013-037}. In
such a scenario, the new method presented is 30\% more sensitive than
a three $q^2$-bin fit to the \CP-averaged, $S_i$, in the region
$1<q^2<6$~\gevgevcccc, improving the discrimination between this
scenario and the SM from 5.0 to 6.5~$\sigma$.  An additional
$\sim$70\% of integrated luminosity would therefore be required in
order for a fit to the \CP-averaged observables to achieve the same
sensitivity as the fit to the \Kstarzb spin-amplitudes. This
improvement arises from the treatment of the $q^2$ dependence of the
amplitudes as a continuous function, and the use of all symmetry
relations of the angular distribution to reduce the number of
independent parameters that describe the \decay{\Bzb}{\Kstarzb\mumu}
decay.

\section*{Acknowledgements}
We would like to thank T.~Blake, S.~Cunliffe and J.~Matias for
insightful discussions on the symmetry transformations of the
spin-amplitudes. We are also grateful to D.~van-Dyk and N.~Serra for
helpful discussions. We also aknowledge support from the Science and
Technology Facilities Council under grant number ST/K001604/1. M.P and
K.P would also like to ackowledge STFC under grant numbers
ST/G005974/1 and ST/K001256/1, respectively.

\clearpage
\clearpage
{\noindent\bf\Large Appendix}
\appendix
\section{Amplitude transformation and parametrisation}
\label{app:a}
The spin-amplitudes in the fixed-basis can be determined by applying
Eq.~(\ref{eq:symm_transf}) to a set of amplitudes in the original
basis. The transformation angles, $\phi_{L},\phi_{R},\theta,\omega$
can be determined in terms of the original amplitudes, by requiring
that in the fixed-basis Eq.~(\ref{eqn:good_cnstr}) holds, and solving
a system of four non-linear equations. The analytical expressions of
the transformation angles are given in Ref.~\cite{quim_swave_sym} and shown below
   \begin{align*}
   \label{eq:trans_angles}
   \tan{2\omega} &=2\frac{\mathrm{Im}(A_{0}^{R})\mathrm{Re}(A_{0}^{L})+(L\leftrightarrow R)}{|A_{0}^{R}|^{2}-|A_{0}^{L}|^{2}}\\
    \tan\theta &=\frac{\mathrm{Re}(A_{0}^{R})+\mathrm{Im}(A_{0}^{L})\tan\omega}{-\mathrm{Re}(A_{0}^{L})+\mathrm{Im}(A_{0}^{R})\tan\omega}\\
     \tan\phi_{L} &=\frac{\mathrm{Im}(A_{0}^{L})+\mathrm{Im}A_{0}^{R}\tan\theta-[\mathrm{Re}(A_{0}^{R})-\mathrm{Re}(A_{0}^{L})\tan\theta]\tan\omega}{-\mathrm{Re}(A_{0}^{L})+\mathrm{Re}(A_{0}^{R})\tan\theta+[\mathrm{Im}(A_{0}^{R})+\mathrm{Im}(A_{0}^{L})\tan\theta]\tan\omega}\\
\tan\phi_{R}&=\frac{\mathrm{Im}(A_{\perp}^{R})+\mathrm{Im}(A_{\perp}^{L})\tan\theta-[\mathrm{Re}(A_{\perp}^{L})-\mathrm{Re}(A_{\perp}^{R})\tan\theta]\tan\omega}{-\mathrm{Re}(A_{\perp}^{R})+\mathrm{Re}(A_{\perp}^{L})\tan\theta+[\mathrm{Im}(A_{\perp}^{L})+\mathrm{Im}(A_{\perp}^{R})\tan\theta]\tan\omega},
\end{align*}    
\noindent where $A_{i}^{L,R}$ are the \Kstarzb amplitudes in the
original basis.

A \texttt{C++} library will be provided shortly, which will transform
a given set of amplitudes to the fixed-basis, and perform a fit in
$q^2$ to determine the coefficients $\alpha,~\beta,~\gamma$ of the
$q^2$ ansatz of Eq.~(\ref{eq:ansatz}).  In addition, provided there is
a consensus in the experimental community, the likelihood function of
Eq.~(\ref{eq:lhood}) could be made publicly available, along with
tools that minimise the likelihood over the nuisance parameters,
for a given set of amplitudes.

\addcontentsline{toc}{section}{References}
\bibliographystyle{LHCb}
\bibliography{main,LHCb-PAPER,LHCb-CONF,LHCb-DP}

\end{document}